\documentstyle[11pt,aaspp4]{article}

\def \halpha {H$\alpha$}
\def \ha {H$\alpha$}

\def\wave#1{$\lambda${#1}}

\def\etal{{\rm et al.}}

\def\secspt{$''_\cdot$}
\def\ltsima{$\; \buildrel < \over \sim \;$}
\def\simlt{\lower.5ex\hbox{\ltsima}}
\def\gtsima{$\; \buildrel > \over \sim \;$}
\def\simgt{\lower.5ex\hbox{\gtsima}}
\def\samename{\vrule height0.4pt depth0.0pt width1.0in \thinspace.}
\def   \flam    {$F_\lambda$}
\def   \pf      {$P\times F_\lambda$}
\def   \qf      {$Q\times F_\lambda$}
\def   \uf      {$U\times F_\lambda$}
\def   \kms     {km s$^{-1}$}
\def   \p       {$P$}
\def   \plam    {$P_\lambda$}
\def   \q       {$Q$}
\def   \u       {$U$}
\def   \pa      {$\theta$}

\slugcomment{Accepted for publication in {\it PASP}, May 1997.}

\begin{document}

\title{Probing the Geometry and Circumstellar Environment of SN~1993J in M81}
\author{Hien D. Tran\altaffilmark{1,~2}, Alexei V. Filippenko\altaffilmark{3},
Gary D. Schmidt\altaffilmark{4}, Karen S. Bjorkman\altaffilmark{5,~6}, 
Buell~T.~Jannuzi\altaffilmark{7}, and Paul S. Smith\altaffilmark{7}}
\altaffiltext{1}{UCO/Lick Observatory, University of California, Santa Cruz,
CA 95064; tran@ucolick.org.}
\altaffiltext{2}{Current Address: Institute of Geophysics and Planetary
Physics, Lawrence Livermore National Laboratory, 7000 East Avenue, P.O. Box
808, L413, Livermore, CA 94550; htran@igpp.llnl.gov.} 
\altaffiltext{3}{Department of Astronomy, and Center for Particle
Astrophysics, University of California, Berkeley, CA 94720-3411; alex@astro.berkeley.edu.}
\altaffiltext{4}{Steward Observatory, University of Arizona, Tucson, AZ 85721;
gschmidt@as.arizona.edu.}
\altaffiltext{5}{Space Astronomy Laboratory, University of Wisconsin, 
Madison, WI 53706.}
\altaffiltext{6}{Current Address: Ritter Observatory, Dept. of Physics and 
Astronomy, University of Toledo, Toledo OH 43606; karen@physics.utoledo.edu.}
\altaffiltext{7}{National Optical Astronomy Observatories, P.O. Box 26732, 
Tucson, AZ 85726-6732; jannuzi@noao.edu, psmith@noao.edu.}

\begin{abstract}

We have monitored the polarized radiation of the Type IIb SN 1993J in
M81 over a period of 41 days, starting from 7 days after the explosion on 
1993 March 27.5 (UT). 
Our data show clear evidence that the intrinsic continuum polarization 
of SN 1993J evolved from being essentially negligible on April 3--4, to a peak 
value of $\sim$~1\% in late April 1993, 
and started to decline by the middle of May. 
The polarized flux spectrum in late April strongly resembled spectra of 
Type Ib supernovae, with prominent He I lines but redshifted 
$\sim$ 3380 \kms~relative to the total flux spectrum.
These data are consistent with models of H\"oflich; they suggest that the 
polarization was most likely produced by either an asymmetric helium core 
configuration of material and/or flux, or scattering from an asymmetric 
circumstellar distribution of dusty material. 
A combination of electron and dust scattering, as well as a
clumpy or stratified distribution of the emitting gas, are 
possible as the polarization mechanism of the continuum and emission lines. 
The latter interpretation is supported by the fact that 1--2 months after 
the explosion, the observed rotations of polarization position angle across 
prominent line features remain even after correction for effects of 
interstellar polarization. This indicates that emission lines of He I, 
Fe II, [O I], and H are all intrinsically polarized at position angles 
different from that of the continuum, with the non-Balmer lines generally 
being most highly polarized.
If the supernova had an oblate
geometry, our data are consistent with a small viewing angle (i.e., more or 
less equator-on), although the degree of asphericity that gave rise to
the polarization at early times is probably smaller (minor to major 
axis ratio $\gtrsim$~0.7) than has been previously suggested.  

\end{abstract}


\section{Introduction}

Supernova 1993J in the nearby galaxy M81 (d = 3.64 Mpc; Freedman \etal~1994) 
was discovered visually by amateur astronomer F. Garcia on 1993
March 28.906 UT (Ripero, Garcia, \& Rodriguez 1993). Early-time data and 
models (see Wheeler \& Filippenko 1996 for a review) suggest a probable
explosion date of March 27.5 (Wheeler \etal~1993; Baron \etal~1993; 
Lewis \etal~1994). Based on the presence of hydrogen in the first few 
spectra, SN 1993J was classified as a Type II supernova 
(e.g., Wheeler \etal~1993). Its visual light curve, however, 
was peculiar: it exhibited a rapid rise to a first maximum around March 30, 
a decline of $\sim$ 6 days duration, a subsequent rise to a second maximum near 
April 18, and a short decline followed by an exponential tail (Lewis 
\etal~1994; Richmond \etal~1994, 1996). The most likely interpretation is that the 
progenitor of SN 1993J was a massive star with only a low-mass outer skin of 
helium-rich hydrogen (e.g., Nomoto \etal~1993; Podsiadlowski \etal~1993). 
The appearance of prominent He I lines $\sim$ 1 month after the explosion, 
together with the diminishing relative strength of \halpha~emission, 
supported this hypothesis and led to a revised classification of Type IIb
(Filippenko, Matheson, \& Ho 1993; Swartz \etal~1993; see also Woosley 
\etal~1987). Late-time spectra (Filippenko, Matheson, \& Barth 1994) 
closely resembled those of SNe Ib, being dominated by [O I] and [Ca II] 
emission, with weak \halpha. On the other hand, \halpha~once again became 
the dominant feature in SN 1993J at $t \gtrsim$~14 months 
(Filippenko \etal~1994; Matheson \etal~1997), most likely due to 
interaction of the ejecta with circumstellar gas.  

Not since the appearance of SN 1987A has there been a supernova subject
to such close scrutiny at all wavelengths, using 
a variety of different observational techniques, including 
polarimetric measurements. 
Spectropolarimetry and photopolarimetry offer a unique view compared to 
other methods; not only do they give information that normal spectroscopy
and photometry offer, but they also provide information about 
the geometric structure of the emitting or scattering regions of the SN
and its environment. This information yields important
clues to understanding the explosion mechanism and expansion of the 
ejecta (Shapiro \& Sutherland 1982; H\"oflich 1995; H\"oflich \etal~1996). 

Due to the brevity and scarcity of bright, nearby SNe, 
polarimetric observations of these objects have been rare. 
Broad-band polarimetry and spectropolarimetry of SN 1987A led to the
conclusion that its ejecta possessed asymmetry of $\sim$~10\% or more
(Cropper \etal~1988; Mendez \etal~1988; Barrett 1988; Jeffery 1991), 
raising questions about the normal assumption that SN 
explosions are essentially spherically symmetric. 
Since this assumption can have important consequences regarding the 
evolution of SNe and their use as calibratable 
candles in measuring cosmological distances (Shapiro \& Sutherland 1982; 
McCall 1984), it is important that we further investigate the true geometry 
of SN explosions. 

Current indications regarding the symmetry of the ejecta of SN 1993J
are contradictory.
Previous spectropolarimetric measurements of SN 1993J 
(Trammel, Hines, \& Wheeler 1993, hereafter THW; Doroshenko, Efimov, \&
Shakhovskoi 1995) have shown that it is significantly polarized.
An aspherical distribution of the 
emission-line regions of SN 1993J has also been suggested based on 
the observed asymmetries of the lines (Spyromilio 1994; Lewis \etal~1994). 
On the other hand, radio observations 
by Marcaide \etal~(1995a, b) have shown that the ejecta envelope of SN 1993J 
appears to be quite spherically symmetric, albeit with variations in brightness
around the shell indicative of asymmetric emission regions. 
However, these sets of data were taken at different times
or ``phases" and thus are not appropriate for detailed intercomparison. 
Moreover, the radio observations and the polarimetry may well probe 
physically different regions of the envelope.
As pointed out by H\"oflich (1995), temporally resolved polarimetric 
information is important in understanding the explosion geometry and 
can provide a crucial discriminant between different theoretical models.

In this paper, we present polarimetric observations of SN 1993J at 
ten different
epochs spanning a period of approximately 41 days, from just 7 days
after explosion to about 28 days after the supernova reached second 
maximum brightness. The good temporal coverage
offers an opportunity never before available: to investigate the complex
configuration (e.g., single or binary system) and explosion geometry 
of a supernova and how it evolves. 
In \S 2 we describe the observations. Section 3 discusses the data 
analysis and results with special attention to the
determination and effect of the interstellar polarization (ISpol), which 
contaminates the true polarization from SN 1993J. In \S 4 we
discuss the implications of these data as interpreted from 
various models of polarization arising from SN explosions. 
We restrict ourselves
to a somewhat qualitative interpretation of the data in terms of existing 
models; a more rigorous examination involving theoretical modeling and 
implications of the complete data set is beyond the scope of this paper.
Our conclusions are summarized in \S 5.

\section{Observations}

Polarimetric observations of SN 1993J were made during a
period of 41 days. All but one epoch employed
spectropolarimetry. Table 1 lists the dates of observations (in all cases UT), 
telescopes and instruments used, as well as the wavelength coverage of each 
observation. Our first observations were made on 1993 April 3 and 4 at 
the University of Wisconsin Pine Bluff Observatory (PBO), with the 
0.9 m telescope and HPOL spectropolarimeter (Nordsieck et al. 1992),
only 7 days after the explosion, and within a day before the first optical 
minimum (Schmidt \etal~1993; Lewis \etal~1994; Richmond \etal~1994, 1996). 
The next observation was made at the Steward Observatory 
1.0 m telescope using the Two-Holer polarimeter (see Smith \etal~1992), 
through a 7\secspt4 circular aperture and broad-band $V$ filter
by P. Smith (see Jannuzi \etal~1993). This observation, obtained about 11 
days after explosion, 
caught the SN on the rise to a second maximum. Subsequent observations were 
made at Kitt Peak National Observatory (KPNO), Steward, and Lick Observatories,
all using spectropolarimetry. The last date for which we have polarimetry 
data is 1993 May 14 when the SN was well on its way to the exponential 
tail of the light curve. Our data set covers a reasonable part of
the temporal history of the light curve, with points surrounding the 
second maximum as well as the first minimum. 
The wavelength coverage ranges from the atmospheric 
cutoff at $\sim$ 3200 \AA~up to 8000 \AA~at some epochs.

At the time of the PBO observations, the HPOL instrument used a Reticon detector,
providing polarimetric spectral coverage from about 3400 to 7600 \AA~with 
a polarimetric spectral resolution of $\sim$ 20 \AA. 
The observations were made using a 12\arcsec~$\times$ 12\arcsec~aperture 
in order to compensate for the atmospheric dispersion over the wide
spectral range.  The data were reduced using standard calibration
and reduction techniques developed specifically for HPOL.  For
details on the instrument design and data reduction process, see
Wolff, Nordsieck, \& Nook (1996).  Due to the faintness of SN 1993J
for the relatively small telescope aperture at PBO, and to the low
polarization at the time of the observation, the signal-to-noise ratio (S/N)
was not adequate to resolve any detail in the polarization spectra.
However, we have combined the two nights of observations together and
binned over the entire spectral range in order to obtain a
measurement of the polarization of the supernova early in its
history.

At KPNO and Steward Observatory (SO), spectropolarimetric observations were 
made at the 4 m Mayall and 2.3 m Bok telescopes, respectively, with the CCD 
spectropolarimeter described by Schmidt, Stockman, \& Smith (1992). 
Either a 3\arcsec~or 4\arcsec~wide slit was used at the SO telescope 
together with a 600 grooves mm$^{-1}$ grating in first order. 
The same grating was used at the 4 m telescope, with the instrumental 
setup and data acquisition being identical to that described by
Smith et al. (1994).

Observations at Lick Observatory were made with the 3 m Shane telescope and
the Kast double spectrograph (Miller \& Stone 1993). For most observations, 
a dichroic splitting the
light at 4600 \AA~was employed with a 600 grooves mm$^{-1}$ grism on the blue 
side and a 600 grooves mm$^{-1}$ grating on the red side. 
A long, 2\secspt4 wide slit was used, oriented near the parallactic angle
to minimize light loss due to atmospheric dispersion.
Thin 400$\times$1200 pixel CCDs were used as the detectors on both sides. 
This setup gives a resolution of $\sim$ 6 \AA~for both blue and red sides, 
corresponding to a dispersion of 1.85 \AA~pixel$^{-1}$ and 
2.36 \AA~pixel$^{-1}$, respectively. 
On April 20, a 5500 \AA~dichroic was used with a 1200 grooves mm$^{-1}$
grating on the red side, giving a dispersion of 1.17 \AA~pixel$^{-1}$.
Data were reduced using the standard VISTA procedures for spectropolarimetry 
as described, for example, by Tran (1995a). 

\section{Results and Analysis}

Figure 1 shows the total flux spectra of SN 1993J at different epochs 
for which we have spectropolarimetric data. They illustrate the visible 
changes in the optical spectra as the SN evolved.
In Table 2, we present a summary of the polarization averaged over the 
wavelength range 4800--6800 \AA, similar to, but somewhat redder than that 
covered by photometry in the $V$ band. 
The results listed for April 3 are the average 
measurements over two nights at PBO.
Table 2 lists both the observed polarizations and those corrected 
for the interstellar contribution. A discussion of the ISpol component
is given below and in Appendix A.
Throughout this paper, in presenting results and displaying data 
we prefer to quote the normalized Stokes parameters \q~and \u, and 
to show plots of the ``rotated Stokes parameter" (RSP) and Stokes flux 
(S $\times$ \flam) rather than \p~and \pf~to avoid statistical 
biases (e.g., Tran 1995b; see also Wardle \& Kronberg 1974; 
Simmons \& Stewart 1985). 
Thus, in this paper we use \p~and \pf~interchangeably
with RSP and S $\times$ \flam, respectively.

\subsection{Interstellar Polarization Component}

Because the effects that ISpol can have on the true polarization of an object 
can potentially influence the interpretation, its
determination is one of the most crucial steps in our effort to understand 
the polarization of SN 1993J. 
When we speak of ISpol we mean all polarization signatures {\it not} 
intrinsically attributable to the source itself. 
In principle, the ISpol component consists of two components, one originating 
from the
interstellar medium (ISM) of our Galaxy, and the other from that of M81, 
the host galaxy. Polarization measurements of several stars near the line 
of sight to the SN have confirmed that the local interstellar component 
within our Galaxy is likely to be small ($\lesssim$~0.1\%; Jannuzi \etal~1993).

The ISpol component to SN 1993J has been discussed by THW, who
assumed the \ha~emission line 
to be intrinsically unpolarized at the time of their observation, 
and thus used the observed polarization of the line as the ISpol
component. Since our data taken on April 20 UT (Fig. 2) were at exactly 
the same epoch as the THW data, they give us an opportunity to compare 
directly with the THW results and to present an independent check of this 
important step. 
We include in Appendix A a complete description of how the ISpol 
component is derived from our data. 
Below, we discuss how it compares with the result determined by THW, 
and what implications it may have on the observations.

Following THW, we also assume that the intrinsic polarization of the 
\ha~emission line on April 20 is zero and
estimate ISpol from the observed polarization of this line. No assumptions
about the polarization of the continuum or line emission at any other epochs
are made. We obtain essentially the same degree of polarization 
as do THW for the {\it observed} continuum (which presumably contains both the 
intrinsic continuum polarization {\it plus} ISpol), but a substantially smaller 
polarization at a different position angle for the \halpha~emission 
[$P$ = 0.63\%, \pa~= 171\arcdeg~compared to 
$P$ = 1.1\%, \pa~= 150\arcdeg~(THW)]. 
Under the assumption that \halpha~is 
intrinsically unpolarized, what is measured is all ISpol. 
This means that the continuum 
\p~after correction for ISpol is substantially less than what THW 
derived (only $\sim$~1\% instead of 1.6\%), and this may have important 
consequences for the physical interpretation of the origin of the polarization, 
geometry of the envelope and/or ejecta, and evolution of the polarization 
(see \S 4).

In Figure 3, we illustrate the spectropolarimetry of April 20 after 
correction for our value of ISpol. 
The procedure for doing this is similar to that
described by THW and Tran (1995a). 
As can be seen, our results show that the ISpol-corrected PA 
flattens out and the rotation across \ha~essentially disappears, providing
some validity to the derived ISpol. 
Significant rotation may still be present in the blue, however.
There is also a curious polarized ``feature" that appears near 4650 \AA~in the 
ISpol-corrected data but {\it not} in the uncorrected data 
(compare Figs. 2 and 3). The significance of this local
polarization maximum is unclear, but we note that our corrected polarization 
spectrum looks very similar to the model calculations of 
H\"oflich \etal~(1996), including the appearance of a similar polarization
hump at roughly the same wavelength. If the interpretation of H\"oflich \etal~is
correct, this feature is actually a result of depolarization effects of
the surrounding combination of lines.
In this regard, the similarity of the corrected polarization spectrum 
to theoretical models lends some credence to the ISpol value derived by us.
The apparent rise of \pf~toward the blue in the red setting 
(i.e., \wave~= 6200 -- 5900 \AA) may signify the presence of polarized
He I + Na I emission that appears stronger in later data (see \S 4.2). 

Another argument further supporting our ISpol
value is that after correction, 
it seems to remove the \halpha~emission line in \pf~more completely than
was done by THW. For comparison, we show in 
Figure 4 the \p~and \pf~spectra (see also Fig. 3 of THW) after 
correction for the ISpol derived by THW. Note that
the use of the THW ISpol produces a corrected \p~of order 1\% higher than that
obtained in Figure 3, and still leaves some residual emission at \ha, which 
by assumption must disappear. 
Thus, all of these characteristics strongly suggest
that the entire \ha~polarization is due to ISpol as correctly assumed by
THW, but our derived value is probably more accurate. 
As also discussed \S 4.1, the very low corrected polarization at early times 
(e.g., April 3) further strengthens the validity of our ISpol value.
Since the $Q$ and $U$ values derived from the continuum subtraction method
were able to accomplish this better than those from the spectral fitting
method (see Appendix A), we prefer the former and adopt 
$P$ = 0.63\%~and $\theta$ = 171\arcdeg~as the ISpol throughout the 
discussion in this paper. 

Although our relatively low value for ISpol $P$ may seem to be in 
conflict with some of the highest reddening estimates derived for SN 1993J, 
it is not inconsistent with the wide range of 
$E(B-V)$ reported in the literature (0.1 -- 0.8 mag; Clocchiatti \etal~1995). 
From the relation $P_{max}<9 E(B-V)$\% (Serkowski, Mathewson, \& Ford 1975), 
we might expect the maximum ISpol $P$
to be anywhere from 0.9\% to 7\%. It should be noted that since this 
relation represents the {\it maximum} polarization expected due to 
dichroic absorption by dust grains in our own Galaxy, the actual 
polarization can be significantly below this upper envelope.

\subsection{Empirical Modeling of Polarization Spectra at Later Epochs}

Since we have data at different epochs, it would be desirable to
isolate the ISpol component at later times. In principle this should be 
possible since the ISpol component must remain unchanged with time.
If such a non-varying component could be identified from the late April and 
mid-May data, it would provide an independent check on our ISpol value 
derived above. In reality, the complexity and changing behavior of the
late-time spectra make such decomposition very difficult. In addition,
this way of extracting the ISpol component is valid only so long as the
polarization of any of the individual components that comprise the flux
spectrum also remained constant with time. 
If they all varied, but we had no information on {\it how} they 
varied, it would not be possible to derive the non-varying ISpol component
since it is coupled with each of them. 
If we suppose that the \ha~emission remains intrinsically unpolarized 
(as we assumed on April 20), we could measure its polarization at
later epochs and see if it matches what we obtained for April 20.
We have made an attempt to extract such a component from the 
observations of April 30 and May 11, representatives of the ``middle" and
``late" phases in our data.
The observed spectropolarimetry for these epochs is shown in Figures
5 and 6, respectively. Although taken only 11 days apart, these data
show that there are significant differences in the spectra and polarizations
between the two epochs. Some of these differences are due to the lower
S/N of the May 11 data, but they can also be attributed to actual 
changes in flux and polarization of various spectral features, as we discuss 
below.

In order to estimate where the polarization arises, it is necessary
to determine the individual components that make up the flux spectrum \flam. 
At these times, it is reasonable to assume that \flam~was composed chiefly of 
{\it at least} three components: (1) a continuum, (2) a hydrogen-rich 
emission/absorption line component typical of most SNe II spectra 
and early spectra of SN 1993J, and (3) a helium-rich and hydrogen-free 
emission/absorption line component typical of SNe Ib 
(e.g., Harkness \& Wheeler 1990) that 
started to show up perhaps very shortly after
April 20, became clearly apparent on April 26 (Fig. 1), and continued to get 
stronger with time (Filippenko \etal~1993; Swartz \etal~1993). 
For the hydrogen-rich component, we continue to use
the April 15 spectrum of SN 1993J that was adopted to fit the April 20
data (see Appendix A). For the helium-rich component, we choose the spectrum 
of the prototypical
SN Ib 1984L obtained 3 weeks after maximum brightness by Harkness 
\etal~(1987). For simplicity, the continuum is chosen to be a constant
across \ha~emission. This level is an approximate estimate of the continuum 
drawn underneath \ha~in the observed \flam~spectrum. 
Since we limit our fit only to 
the spectral region containing \ha~emission, our choice of a constant 
continuum is not expected to be significantly in error. 

As described in Appendix A, least-squares fitting is performed first on the
total flux spectrum \flam, to determine the relative contribution of each
component; the appropriately scaled constituents are subsequently used 
to fit the Stokes spectra \qf~and \uf~to obtain the \q~and \u~parameters.
It is in this last step that we recognize that a fourth component is 
needed to improve the fit. Examination of the \qf~spectrum shown in 
Figure 7 reveals that it is remarkably similar to the total flux spectrum
of a SN Ib but redshifted by $\sim$ 3380 \kms. 
Also illustrated in Figure 7 are the individual H-rich and redshifted He-rich 
line components used to fit the total flux spectrum. For clarity, the 
unshifted He-rich component is not shown.

Figure 7 demonstrates that the fit to the flux spectrum is reasonably good.
The main discrepancies are that the observed spectrum exhibits narrower,
slightly deeper, and less blueshifted Na I D and \ha~absorption (due to 
the recession of the photosphere into the ejecta), as well as possible
[O I] \wave 6300 emission (an indication of the beginning of the transition 
to the nebular phase). The best fit to \qf~(heavy solid line in Fig. 7c) is 
achieved only with the addition of the fourth, redshifted SN Ib component
(hereafter referred to as the ``He-rich component"). 
Note that for the purpose of illustration, we have extended the fit to 
encompass a larger wavelength range, and the continuum
has been fitted with a third order polynomial (instead of a constant) to 
allow for variation in wavelength. 
In \qf, the resemblance to SNe Ib is striking. 
For example, the P-Cygni profile due to He I \wave 5876 + Na I D appears to 
match that in the SN 1984L spectrum but shifted by $\sim$ 3380 \kms, and 
likewise for the profile of He I \wave 6678.
In fact,
our fit shows that in \qf, virtually all of the flux can be accounted for
by the continuum and this ``redshifted" He-rich component, and very little
comes from the unshifted He-rich and H-rich components, yet these 
represent the bulk of the total flux in the lines. 

With the inclusion of this fourth component, the results of our fit 
constrained around the \ha~region (6390--6860 \AA, as for April 20) 
for \flam, \qf, and \uf~for the April 30 and May 11 data 
are shown in Figures 8 and 9, respectively, and are summarized in Table 3. 
The result is that the ``redshifted" He-rich component, if real, makes a small
contribution to the flux but is responsible for the most of the 
observed polarization of the lines. 
The origin and significance of this redshift are unclear, but it could
be due to light being scattered by the expanding photosphere or by 
dusty circumstellar material not in the photosphere (see \S 4.2).
In general, the fits all appear reasonably good.
These results should, however, be taken with caution because of the following
inadequacies and limitations of the procedure. 
(1) The three components used to fit 
the lines in \flam~are inadequate to fully represent the complexities of the
spectra, as can be seen, for example, 
by the possible presence of [O I] \wave 6300 emission 
blueward of \ha~absorption.
(2) The spectra of SN 1993J (April 15) and SN 1984L may not be 
representative of the line profiles of {\it all} different epochs observed.
(3) No attempt was made to take into account such effects as line
broadening resulting from electron scattering that may have caused the
polarization, or other mechanisms that have reprocessed the scattered 
radiation. 
(4) By design, absorption of light after being scattered (and polarized) 
cannot be treated by our procedure. 
The poor fit around 6400--6500 \AA~in Fig. 8b,c may be due 
to absorption of polarized light by the 
expanding hydrogen and helium gas.
(5) While the ``redshifted" He-rich component is 
empirically justified due to its remarkable resemblance to the \qf~spectrum, 
its addition is rather {\it ad hoc}, chosen to achieve the best fit. 
We do not know if there are any other low-level
components that may have escaped detection. 

Nevertheless, these results
are useful for qualitative illustration and can reveal some important
characteristics.
First, our exercise did not detect any component with polarization 
magnitude and orientation matching the ISpol of THW or this paper.
Specifically, the observed polarization of pure \ha~emission, which we assumed 
to be intrinsically unpolarized on April 20 and thus defined the ISpol, was 
changing at every epoch. Inspection of Tables 3 and 4 also indicates that the 
continuum polarization appeared to vary in 
time, with the general trend of increasing in magnitude between April 20 
and April 30 and declining between April 30 and May 11.
The polarizations of all other components also varied between April 30 
and May 11.
Thus, unless we know exactly how these polarization components changed, 
the ISpol cannot be directly extracted from these observations.
Second, these results demonstrate that the emission lines are 
in fact polarized, especially those associated with the He-rich component,
which if real may be polarized up to $\sim$ 8\%!
This is in contrast to most theoretical expectations that the lines 
are essentially unpolarized due to the assumed complete 
redistribution of photons over the line profile (H\"oflich 1995). 
That the polarization changed quite substantially
between April 20 and April 30, and that the line emission possessed 
significant polarization at later times, will be discussed further in \S 4.2. 
Third, the presence of (possibly very high) polarization in the Type Ib-like 
redshifted line component strongly suggests that whatever is connected with
the onset of He I lines in the late-April spectra of SN 1993J is also
most likely responsible for the surge in polarization. 
This is investigated in more detail below.

\section{Discussion}

\subsection{Temporal Evolution of the Polarization}

Recently, H\"oflich \etal~(1996) have analyzed the polarization of SN 1993J
with theoretical models in an effort to understand and constrain the geometry
of the ejecta and envelope and/or identify the polarization mechanism. One
crucial element in their analysis is the assumed value of interstellar 
polarization, ISpol. H\"oflich \etal~adopted the ISpol of THW, and 
consequently found that the intrinsic polarization of SN 1993J was 
surprisingly high at the earliest epoch of our observations, only 
$\sim$ 7 days after the explosion. Furthermore, the polarization seemed 
to remain constant in both magnitude and direction through late April.  
The data of Doroshenko \etal~(1995) appear to have confirmed this 
early trend, which remained until the middle of May; however, they have 
a gap in coverage between the third week of April through the first week 
of May, and any possible polarization changes during this period would 
have been missed.
As discussed by H\"oflich \etal, the observed polarization behavior 
from those data is unexpected and difficult to 
reconcile with current theoretical models. 
In particular, we should expect the initial polarization (early April)
to be small when the photosphere is optically thick and still 
in the outer portion of the hydrogen envelope, and it should rise 
at later times when the photosphere has receded to the top of the He core
and the envelope has become more optically thin. 

We can make some comparison
and evaluation of the current data based on the published models of 
H\"oflich \etal~These models examine the data in terms of the line profiles in both
total flux spectrum \flam~and polarization spectrum $P_\lambda$. Important
parameters to consider are the breadth and depth of the absorption
components, the line shift in both \flam~and \plam, as well as the 
amount of depolarization in the line. Consideration of these parameters
should give us constraints on the structural and geometric parameters such
as minor to major axis ratio of the envelope $E$, density 
gradient $n$ of the atmosphere,
and inclination of viewing angle $i$. Prolate or oblate geometry can also be
evaluated and constrained, in principle. Because of the higher assumed ISpol
$P$ in the H\"oflich \etal~analysis, the absolute polarization after correction 
is about $\sim$ 1\% higher than that seen in our results.
Given these differences, it is conceivable that for the 
\flam~and our corrected polarization spectrum of April 20 shown 
in Figure 3, a reasonable fit would require that we keep the 
inclination large (as concluded by H\"oflich \etal~1996) to match the 
width and blueshift of the total flux absorption line profile, but possibly 
decrease the asphericity (i.e., increase the axis ratio $E$) in order to 
match the lower \p. 
H\"oflich \etal~(1996) found that an axis ratio of $E$ = 0.6 best matches 
the data of THW. Our data would probably require $E \gtrsim$~0.7. 
For comparison, the models of Shapiro \& Sutherland (1982) suggest
that, with an oblate geometry and equatorial viewing, 
the maximum intrinsic polarization of $\sim$~1\% in our data would imply 
an axis ratio of $E$ = 0.76 for cases with absorption in addition to
scattering, and $E$ = 0.47 for cases without absorption.
Since the breadth of the polarization minimum in \pf~is about the same for
our correction and the H\"oflich \etal~result, the density gradient would
probably remain close to $n$ $\approx$ 5 as deduced by H\"oflich \etal~Thus, 
the main effect of the smaller corrected polarization derived 
in this paper (as a result of the smaller derived ISpol) is to 
reduce the asphericity of the envelope. 
A far more significant consequence, however, is on the temporal evolution of 
the polarization.

In Figure 10 we show in (\q, \u) space the flux-weighted average polarizations 
listed in Table 2 both before and after correction for the ISpol. 
Also shown in star and triangle symbols are the ISpol derived from 
this paper and THW, respectively. The line connecting each point
in chronological order starting with the first epoch April 3 (lower right)
illustrates how the polarization magnitude and PA have varied. 
Note that the effect of removing the ISpol is to shift the entire set
of points to the upper left quadrant. 
We also show the corrected polarizations of Table 2 in a different form 
in Figure 11. Here the $V$-band polarizations and position angles are shown 
as functions of time in days after the assumed date of explosion, 
1993 March 27.5 UT. A light curve in observed $V$ magnitudes 
from Lewis \etal~(1994) is also shown for comparison.
We note that although the polarization data in late April 
were obtained at different telescopes and instruments (Table 1), 
their similarity indicates that these measurements are highly consistent with
each other. The formal uncertainties, as calculated from photon statistics, 
of the observed polarizations are fairly small (Fig. 10), but
the uncertainties in the corrected polarizations, which are of 
order 0.1--0.2\%, are dominated by the uncertainties in the inferred ISpol. 

Our first observation at PBO indicates that it measured entirely the 
ISpol component, as evidenced by their similarity in Figure 10. 
Thus, the supernova was intrinsically unpolarized at this time. 
Note that while our derived ISpol value is far from proven, some values 
can be excluded. For example, if $Q$(IS) = 0.5\% and $U$(IS) = 0.5\%, the 
intrinsic polarization at the very beginning would be too high 
($\sim$ 0.7\%) to be consistent with the assumption of zero polarization
at this time.
Our second measurement, obtained using the broad $V$ band on April 7, 
just after the first minimum, initially suggested that the supernova was 
unpolarized, but the ISpol correction revealed that the intrinsic 
polarization was actually beginning to rise. 
Throughout the period early April to May the polarization PA also 
varied by about 40$\arcdeg$. 
These observations show that the polarization clearly evolved: 
it was intrinsically non-existent at early
times (April 3, 4), rose to a maximum value at some later time (April 26-30), 
then started to decline gradually (May 11, 14), accompanied by a change 
in position angle. This is just the type of behavior that would be expected 
according to various models (H\"oflich 1995; H\"oflich \etal~1996).

H\"oflich (1995) describes three different geometries for modeling the 
polarization of SN 1993J. These are (I) ellipsoids throughout the star, (II) an
aspherical inner region surrounded by a spherical shell, and (III) a spherical
envelope configuration illuminated by an off-center source.
The strong asphericity required by the polarization of $\sim$ 1--2\%,
as well as the slow time evolution predicted by the first scenario 
(H\"oflich 1995), make this picture seem unlikely.
Although the viewing inclination and the exact phases of our observations 
relative to model calculations are unknown, the rapid rise in polarization 
shown in Figure 11 
suggests that they are probably more consistent with his model II or III,
both of which predict a significant change of \p~with time.

H\"oflich (1995) dismisses scenario II as the cause of polarization 
of SN 1993J based on the high early polarization and the lack of temporal
variation in the two polarization observations of which he was aware 
at the time. We have shown, however, that the April 3 measurement 
can be explained as being due 
entirely to ISpol, and that indeed the polarization was initially very low
but increased later, perhaps coincident with the 
expansion of the photosphere. 
This behavior is very similar to the temporal evolution model of the 
polarization presented by H\"oflich: with time, the outer envelope 
becomes more transparent and the polarization should rise rapidly and 
change PA.
Thus, according to the simple models of H\"oflich, our data 
appear to support the picture that the explosion in SN 1993J was 
asymmetric at the core but more spherically symmetric in the 
outer envelope, a configuration that could reconcile the presence of 
measurable optical polarization and the apparent 
observed symmetry of the radio images reported by Marcaide \etal~(1995a).
On the other hand, it has been suggested that the progenitor of SN 1993J 
was part of a close binary system (Nomoto \etal~1993; Podsiadlowski 
\etal~1993; Ray \etal~1993; Bartunov \etal~1994; Utrobin 1994; 
Woosley \etal~1994; Suzuki \& Nomoto 1995). This could provide a 
natural explanation for the asymmetric illumination pattern of the 
envelope (due to the residual motion of the neutron star sustained
during explosion), consistent with H\"oflich's model III. 
As suggested by Chugai (1992), such asymmetry could also arise from 
an off-center light source of a radioactive nickel blob. 


\subsection{Polarization of the Emission Lines}

We have discussed the polarization of the \ha~emission line in the April 20
data and the hypothesis of it being due entirely to ISpol toward SN 1993J. 
We now address the possibility of intrinsic polarization in emission 
lines at later epochs.
As presented in \S 3.2, our attempt at separating the continuum and 
line polarization shows that essentially all emission lines, 
including the Balmer lines, appear to be intrinsically 
polarized in the late April and May data. 
That these lines are intrinsically 
polarized (and polarized differently than the continuum) can also be 
supported by two key characteristics.
First, in marked contrast to the data 
of April 20, substantial rotation of polarization PA 
at the wavelengths of several strong lines 
is seen before and even after the ISpol is removed, regardless of the
assumed value of ISpol (THW or our result). 
We show in Figures 12 and 13 the spectropolarimetry of April 30 after
correction for our ISpol value and that of THW, respectively.
As can be seen, there are clear rotations of PA across the He I \wave 5876 +
Na I D profile, the \halpha~+ He I \wave 6678 complex, and around the
[O I] \wave 5577 and \wave 6300 regions, both before (Fig. 5) and after 
ISpol correction. 
It is difficult to conceive of a single ISpol value that can completely 
remove these rotations. 
Since rotation in polarization PA with wavelength most likely signifies
the presence of multiple, differently polarized components,
this shows that some of the lines are intrinsically polarized. 
Unless the polarization mechanism in the continuum itself causes these 
effects, which seems unlikely, this represents the single most important 
and visually apparent evidence for the intrinsic polarization in the lines. 

The second argument for intrinsically polarized lines is that changes 
in \p~with time are observed both 
{\it before} and {\it after} ISpol was removed (see Fig. 10). 
If ISpol alone affected the intrinsic polarization, then the 
temporal evolution of $P$ magnitude is not expected. 
Alternatively, these changes could simply be due to relative changes 
in flux of the diluting component from the SN as it brightened or dimmed. 
For example, since the average values of $P$ in Figure 11
were computed as total polarizations containing both continuum and
line flux, if some unpolarized line flux were to vary during this 
period (or if the continuum flux were to vary, simulating changes in line 
flux relative to the continuum), then it could make the total polarization 
vary accordingly. 
However, we would {\it not} expect the observed change of PA (Fig. 11) 
or polarized flux (Figs. 2, 5, 6) with time. 
These changes are an indication that the temporal changes 
in \p~represent a true interaction of evolving line and continuum 
polarizations. 

Note that the Fe II and He I lines seem to be polarized 
more strongly than the Balmer lines (Table 3; Figs. 12, 13). 
The presence of different line polarizations can be seen by examining 
the \pf~spectra (Figs. 12, 13): there are maximum and minimum features 
resembling P-Cygni profiles of Fe II, He I, and perhaps [O I] emission. 
These lines, especially He I, correspond to the Type Ib-like component 
discussed in \S 3.2. Their presence in the polarized flux spectra indicates 
that they are intrinsically polarized. Examination
of the two \pf~spectra corresponding to two different ISpol values also 
shows that there are some slight differences between 
them, such as the prominence, shape, and velocity shifts of the
absorption/emission profiles. More detailed theoretical 
analysis like that of H\"oflich \etal~(1996) could 
provide some insights into the significance of these features. 

As discussed in the previous section, 
our data could be consistent with model II of H\"oflich (1995),
in which the core is more asymmetric than the outer envelope. One
way that we can account for the different polarizations of the lines is 
if they do not arise in the same spatial regions of the SN ejecta. 
According to this picture, our data would suggest that the
Fe II and He I lines come from an asymmetric, deeper region in the SN than 
the hydrogen Balmer lines. 
In other words, our observations may imply a configuration in which
the supernova has a more distorted helium core buried within a less distorted 
envelope. This may also help explain the apparent symmetry of the
radio images observed by Marcaide \etal~(1995a), since they probe
more or less the outer envelope of the supernova.
The ``stratification" in line polarizations implies that the emitting gas 
from the supernova ejecta is compositionally stratified, a 
conclusion also suggested by Swartz \etal~(1993) based on the absence of 
strong contamination of metal lines like [Ca II] and [O I] in 
the observed spectra, and on comparison with their mixed and unmixed models. 
However, the very high polarizations derived for the lines (see \S 3.2),
if real, would make it difficult for them to arise solely from the asymmetric 
geometry of the envelope; highly asymmetric structure would be required.

The redshifted polarized lines (Figs. 7, 12) could naturally arise
from electron scattering of line radiation by the homologously 
expanding photosphere (Jeffery 1991). 
The presence of polarization also suggests that the line emitting
regions may have a clumpy and asymmetric gas distribution.
A non-uniform line emitting pattern can arise from shock emission 
as the ejecta interact with a clumpy or asymmetric distribution of 
circumstellar medium, or from Rayleigh-Taylor instabilities in the cool, dense
shell of gas behind the reverse shock (Chevalier, Blondin, \& Emmering 1992). 
Evidence for clumpiness in the interaction region comes from X-ray 
observations (Suzuki \& Nomoto 1995), from the observed asymmetries 
and blueshifts of emission-line profiles (Lewis \etal~1994; Spyromilio 1994; 
Filippenko \etal~1994), and from early-time radio observations
(Van Dyk \etal~1994).
Also, as mentioned previously, late-time radio observations by 
Marcaide \etal~(1995b) have shown that the ejecta are relatively symmetric, 
but there is an asymmetric emission pattern in the outer envelope, suggesting
perhaps a non-uniform distribution of the circumstellar medium.
Houck \& Fransson (1995), however, have dismissed large-scale asymmetry
in the ejecta as a likely explanation for the blueshift
of [O I] \wave 5577 observed by Lewis \etal~(1994) and
Spyromilio (1994), but instead prefer to interpret it
as being due to blending with iron and cobalt emission lines.
The presence of polarization in the [O I] line
tends to favor the asymmetry interpretation; moreover, Matheson \etal~(1997)
find that [O I] \wave 6300 has essentially the same profile as
[O I] \wave 5577, in apparent conflict with the Houck \& Fransson hypothesis.

An alternative way of producing the high 
polarization is from scattering by circumstellar dusty material,
the so-called ``light-echo" effect (Wang \& Wheeler 1996), which could
also produce a redshifted polarized spectrum, as observed in a planetary
nebula (Walsh \& Clegg 1994). 
Dust grains that survived the initial blast of ultraviolet light
could be present and effectively scatter both continuum and line 
radiation coming from the supernova. 
The PA rotation in the lines could then arise from line emission regions
distributed in a geometry that is different from that of the continuum.
In this case,
an additional polarization signature due to dust scattering, such as the 
wavelength dependence of polarization and polarized flux, might also be 
expected. 
It is difficult to predict what type of wavelength dependence dust scattering
may impose on the polarization without knowing the characteristics of 
the scattering material, but small dust grains, which have high polarization
efficiency, generally produce \p~that rises steeply to the blue because
of the rapidly increasing scattering cross section with increasing
frequency. However, 
a conspiracy of a diluting continuum that is dropping to the blue
or a pure electron scattering atmosphere with a mixture of both 
scattering and absorption (Shapiro \& Sutherland 1982) can also produce 
similarly blueward rising wavelength dependence on \p. In any case, 
there is little evidence of such wavelength dependence in our data, although
depolarization effects by a number of lines in the spectra combined with
decreasing S/N toward the blue make it difficult to be certain.
Again, careful quantitative investigation of the full
data set (our data plus those of THW and Doroshenko \etal~1995) 
in comparison with theoretical models such as those 
of H\"oflich (1995), H\"oflich \etal~(1996), and Wang \& Wheeler (1996) 
should be done to better evaluate different possibilities and 
constrain various parameters.

\section{Summary and Conclusions}

THW and this study have come to different conclusions for 
the ISpol toward SN 1993J. It is worthwhile to reexamine them 
and evaluate their merits as well as the effects that they have on the 
interpretation of the data.
Fortunately, the two cases of ISpol do result in significantly different 
consequences that can, in principle, be distinguished by our current 
understanding of supernova atmospheres and explosion mechanisms. 
First, comparison of the ISpol-corrected \pf~spectra 
shows that \ha~emission is essentially absent in our case but not in THW. 
Although \ha~emission may not be completely unpolarized 
due to effects of Thomson scattering (H\"oflich \etal~1996), by definition
and assumption the line flux {\it must} go to zero in the ISpol-corrected
polarized flux spectrum when the line polarization is properly determined.
Second, the two possibilities of ISpol predict different evolution of the SN
polarization. Application of our ISpol indicates that at $t \approx$ 0,
\p~$\approx$ 0\%, and as time proceeded the polarization increased steadily, 
reached a maximum, and started to decline in mid May. Meanwhile,
the polarization PA rotated from about $\sim$ 80\arcdeg~to $\sim$ 45\arcdeg.
This behavior is remarkably similar to what one would expect
from models II and III of H\"oflich (1995).
On the other hand, the use of the THW ISpol value would predict that the 
polarization was already remarkably high ($\sim$ 1\%) at early times, and 
remained more or less constant in both magnitude and orientation 
throughout the time between early and late April, when SN 1993J 
underwent drastic changes in its spectrum and in the physical conditions 
of its envelope (H\"oflich \etal~1996). 
It is difficult to understand how the polarization
could already be so high in early April, when the envelope was still
optically thick, and remain essentially the same in late April, when it was
optically thin and the photosphere mostly likely had receded into the 
He mantle, giving rise to the He lines. 
Although we have not proven that the ISpol value derived in this paper
is definitively the correct value, 
these two differences perhaps represent the best argument for its validity.

In general, the polarization can be due to (a) a true anisotropy 
in the shape of the ejecta with a uniform distribution of flux, 
(b) intrinsic core asymmetry in the explosion, (c) an asymmetric 
distribution in flux of a radiation source in spherically symmetric 
ejecta, or (d) scattered radiation from dusty circumstellar material.
The temporal evolution seen in our polarization 
observations tends to support the hypothesis of either an 
asymmetric configuration in the supernova core, or an asymmetric flux
distribution, perhaps in the form of an explosion of a binary system 
or an off-center radioactive source. 
Additionally, it may imply that the ejecta interacted with  
clumpy and anisotropic circumstellar material, with dust scattering 
possibly playing a role at later times.

One of the most interesting and surprising results of this paper 
is that, unlike the Balmer lines at early times, 
the emission lines of He I, Fe II, [O I], and H roughly 1--2 months after 
the explosion are all intrinsically polarized at position angles different 
from that of the continuum, with the non-Balmer lines generally being 
most highly polarized. This suggests that the line emission regions 
probably either have a highly anisotropic flux distribution, or are 
stratified in spherically and differentially asymmetric ejecta. 
Full theoretical treatment of the complete data set with the 
revised ISpol adopted in this paper is needed to gain more quantitative
constraints on all of the relevant parameters describing the structure and
geometry of the supernova envelope and ejecta.

\acknowledgments

We are grateful to Luis C. Ho, Andr\'e Martel, and Tom Matheson for 
assistance with some of the observations at Lick Observatory, and to 
Richard Elston for 
assistance with the KPNO observations. Much support was provided by 
the staff members of Pine Bluff, Steward, Lick, and Kitt Peak National 
Observatories. We thank Craig Wheeler and Lifan Wang for instructive 
discussions. A.V.F's research is funded by National Science Foundation
grant AST-9417213. This paper was completed while A.V.F. held an appointment
as a Miller Research Professor in the Miller Institute for Basic Research
in Science (U.C. Berkeley). During the course of this work,
H.D.T. was supported by postdoctoral research fellowships at the California 
Institute of Technology through NSF grant AST-9121889, and at Lick Observatory 
through grant AST-8818925. G.D.S. acknowledges support from NSF grant
AST 91-14087 and P.S.S.'s research is funded by NASA grant 
NAG 5-1630. K.S.B. and spectropolarimetric observations at PBO have been 
supported under NASA contract NAS5-26777 with the University of Wisconsin.

\appendix
\section{Determination of Interstellar Polarization toward SN 1993J}

We derive the ISpol toward SN 1993J 
by extracting the polarization of \ha~emission from the observed 
data of April 20, under the assumption that the intrinsic polarization of
this line is zero at this time. 
Separation of the continuum and line polarization is generally achieved 
by two methods. In the first, the component ``Stokes parameter fluxes" 
(\qf~and \uf) are computed; the best continuum is subsequently drawn under 
the emission line of interest and 
removed. The same task is performed on the total flux spectrum \flam.  
The \q~and \u~of the line alone are the ratios of the 
continuum-subtracted line fluxes of the Stokes flux to the total flux
spectra:

$Q_{line} = Q\times F_\lambda (cs)/F_\lambda (cs)$,

$U_{line} = U\times F_\lambda (cs)/F_\lambda (cs)$,

\noindent
where ``cs" denotes continuum-subtracted line fluxes.

The major uncertainty in this method comes from the placement of 
the continua in both the \q~and \u~fluxes and in the total flux spectra, 
especially the former because of their inherently noisier nature.
Alternatively, estimation of the line and continuum polarizations
can be accomplished by means of a least-squares linear fit of
at least two components, continuum and emission line, to the Stokes 
spectra. In this procedure, the (constant) coefficients to be fitted 
are the Stokes parameters (\q~and \u) of each component, and the 
individual model components are the continuum and line profiles found 
to best match the total observed continuum and line profile in \flam: 

$Q\times F_\lambda(obs) = Q_{cont}F_{cont} + Q_{L1}F_{L1} + Q_{L2}F_{L2} + ...$

\noindent
and similarly for \uf, where

$F_\lambda(obs) = F_{cont} + F_{L1} + F_{L2} + ...$~.

\noindent
Here, ``obs" represents the observed quantities, and the subscripts $cont$ and 
$L1$, $L2$, ... denote the continuum and individual line components, 
respectively. 

This method requires that we initially choose the appropriate forms 
for the continuum and individual ``template" line profiles that can serve
as models in the fit. The individual components used for fitting \qf~and
\uf~are determined by fitting the total line flux profile in \flam.
This method has an advantage over the first in that one does not 
need to guess a continuum in \qf~and \uf. 
It should be limited to a sufficiently small wavelength range, however,
to ensure that the wavelength dependence of polarization, if any, can be
neglected in the fit.
In principle, with good S/N data, the two
methods should give similar results, provided that the models in the
second method are appropriately chosen.

In practice, these methods have difficulties with the
very complex late-time spectra of SN 1993J, which are
superpositions of many different components of both 
emission and absorption lines coming from a variety of different 
atomic species (for example, the appearance of He I absorption 
atop \ha~emission; Filippenko \etal~1993; Swartz \etal~1993).
The difficulties in drawing the ``correct" continuum and knowing the
true line profiles of the various lines present in the late-time spectra
prevent us from successfully disentangling the polarizations
of all possible line components. The combined observed polarizations, 
however, are still very informative; we discuss our attempt
at the decomposition together with implications of these results in \S 3.2.

In terms of complexities in spectral features, the relatively early 
spectropolarimetric data of April 20 (Fig. 2) are most easily interpreted.
The \flam~spectrum resembled those typical of ``pure" SNe II, showing broad 
\halpha~emission and absorption without noticeable contribution from the He I 
lines that appear in later spectra. 
These characteristics and the good S/N allow a 
relatively simple application of the two techniques described 
above for separating the polarizations of the \ha~line and continuum. 
For the least-squares fitting method, we choose to use 
an earlier spectrum taken on April 15 (Filippenko \etal~1993) as 
a template for the \ha~profile, and a constant value underneath 
\ha~emission for the continuum. 
Limiting the fit to a narrow wavelength range minimizes errors due to 
any wavelength dependence of the continuum shape, and use of the 
April 15 data helps to ensure that the profile is sufficiently ``pure" 
in hydrogen and free of possible contribution from He I.
In addition, since the two observations were fairly close in time, 
it is reasonable to assume that only the line's intensity and not its shape 
had changed appreciably. This allows us to use the earlier line profile 
to approximate the \ha~strength relative to the continuum for the later data. 

The fit was performed only in the region of \ha~emission, 6390--6860 \AA.
The results for the fit are shown in Figure 14 and listed in Table 4 
along with the corresponding values from THW. 
[Note that the THW $Q$ and $U$ values are derived 
from \p~and position angle (PA) values given by them 
assuming $P=(Q^2+U^2)^{1/2}$ with no corrections for statistical biases.] 
Before we make direct comparisons of the results, it should be noted 
that, as can be seen from the {\it observed} polarization spectra  
in Figure 2 of this paper and Figure 1 of THW, our data on April 20 are 
consistent with THW in the sense that there is no difference 
in the overall level of the {\it observed} polarization $P$ and position angle
\pa. Moreover, we note that these data are also consistent with those obtained
at very nearly the same time (April 19.87 UT; $P$ = 0.66 $\pm$ 0.09, 
\pa~= 20.4 $\pm$ 3.9 in $V$ band) by Doroshenko \etal~(1995).
Note that the continuum and line profile chosen are able to reproduce the
observed flux reasonably well (Fig. 14), and Table 4 shows that the 
polarization results for the two different methods described are consistent 
with each other, providing some reassurance that the values derived are 
not grossly in error.

For the separated continuum and line components, Table 4 shows that 
our $Q$ and $U$ values for the continuum 
alone are consistent with those of THW. 
However, differences appear in the polarization of \ha~alone;
the $Q$ value agrees well with THW, 
but there is a large discrepancy for $U$ by about a factor of 4 between 
the two studies. 
Given the good agreement in the other three quantities, 
the discrepancy could not have come from error in the 
analysis, but most likely arises from noise in the data and assumptions 
used in deriving $U$. 
A clue to the origin of this 
difference comes from looking at \qf~and \uf~separately; 
see Figure 14b, c.
There is a broad minimum at $\sim$ 6400 \AA~which
occurs only in \uf~and not in \qf, 
corresponding to the same feature noted in Figure 3 of THW (who did
not show \qf~and \uf~separately but only the combined ``Stokes Flux" 
S $\times$ \flam).
THW interpreted this minimum as arising from absorption by the hydrogen 
envelope as the electron-scattered (and polarized) continuum passes through 
it. If so, the minimum should have no effect on the polarization of the 
emission line (or the continuum for that matter). Since it
is blueshifted away from \halpha~emission, this minimum should contribute 
little to the $U$ value of \halpha~emission. Hence, we derived a low number 
for \u(\halpha). 
If, however, the dip is interpreted as being due to blueshifted 
\halpha~in {\it emission} that is intrinsically polarized in the sense opposite
from that of the surrounding continuum, then one recovers a high $U$ value. 
If the flux spectrum is blueshifted by $\sim$ 6590 km s$^{-1}$ 
so that the \halpha~emission coincides
with the dip in \uf, we obtain \u(\ha) = $-$0.96\%, matching
the THW value. However, this same blueshifted line profile
cannot give $Q$ = 0.6\% which both THW and we derive, but underestimates
it by about a factor of 2. 
It is unlikely that the $U$ value of THW was obtained in this
manner, for the spectral shift needed is much too large to arise
from error in the data reduction alone. 
The source of the discrepancy in $U$ is more likely due to 
a combination of the somewhat lower S/N in THW data, the uncertainty in the 
placement of the continuum level, and the difference in wavelength range 
over which \halpha~polarization was determined.
Since one clearly cannot use the observed line flux 
to derive one Stokes parameter and blueshift it to obtain the other 
(the line is either blueshifted in both \q~and \u~fluxes or not at all), and 
since blueshifting the emission line is inconsistent with the interpretation 
that the minimum at 6400 \AA~is due to absorption of the scattered continuum, 
we adopt the values of \q~and \u~derived with the observed, unshifted line 
profiles as ISpol toward SN 1993J. 

How good is the assumption that the \ha~polarization is entirely due to ISpol? 
As mentioned by THW, the two features that make this interpretation 
compelling are (1) the polarization PA is parallel to the spiral arms and 
magnetic fields of M81, and (2) after correcting for this ISpol, the continuum 
PA flattens out and the PA rotation in \halpha~disappears.
Note that although our polarization and PA for \halpha~are considerably 
different from the THW values, they are {\it not} inconsistent with either of 
these points. 
Examination of the M81 magnetic field maps of Beck, Klein, \& 
Krause (1985) shows that given the uncertainties in the map, in
THW, and in our measurements, both PAs are consistent with being parallel to 
the spiral arms and magnetic fields. 
The fact that {\it both} ISpol values are able to result in the flattening 
of PA across \halpha~is probably due to a combination of noise, which may 
hide any variations, and a subtle effect: they both correct for the 
polarization in the right direction (albeit with different magnitudes).
We know that in the continuum, $U$ is positive and $>$ $Q$, which is $\simlt$ 0.
Thus, the continuum PA lies in the range 45--55\arcdeg. In \halpha~emission,
since $Q$(ISpol) = +0.6\%, the corrected $Q$ tends to be small and negative. 
Therefore, if $U$(ISpol) is negative, which is true 
for both THW and our ISpol, the corrected $U$ in the line remains
positive and PA lies in the range 45--67\arcdeg. 
This is sufficiently close to the continuum PA range that, within the noise 
of the data, any differences would not be detectable.

\clearpage
\begin{deluxetable}{lrcr}
\scriptsize
\tablewidth{0pc}
\tablecaption{Polarimetric Observations of SN 1993J}
\tablehead{
\colhead{UT Date} & \colhead{Telescope}  & 
\colhead{$\lambda$ Coverage (\AA)}  & \colhead{Exp. (s)}}
\startdata
1993 April 03$^a$  & 0.9 m PBO     &  3400--7600 & 6451 \nl
1993 April 04$^a$  & 0.9 m PBO     &  3400--7600 & 6451 \nl
1993 April 07  & 1.0 m Steward &  $V$ band~~~    & 4 $\times$ 240 \nl
1993 April 20  & 3.0 m Lick    &  3200--7300 & 4 $\times$ 180 \nl
1993 April 26  & 4.0 m KPNO    &  4000--7250 & 8 $\times$ ~~20 \nl
1993 April 27  & 2.3 m Steward &  4000--7300 & 8 $\times$ ~~40 \nl
1993 April 28  & 2.3 m Steward &  4000--7300 & 8 $\times$ ~~40 \nl
1993 April 30  & 3.0 m Lick    &  3200--7400 & 4 $\times$ 600 \nl
1993 May   11  & 3.0 m Lick    &  3200--7400 & 4 $\times$ 300 \nl
1993 May   14  & 2.3 m Steward &  4100--7440 & 8 $\times$ ~~80 \nl
\tablenotetext{a}{Measurements from these two nights are averaged in the 
analysis presented here.}

\enddata
\end{deluxetable}

\clearpage
\begin{deluxetable}{lrcrcccc}
\tiny
\tablewidth{37pc}
\tablecaption{Observed and ISPol Corrected Polarizations of SN 1993J$^a$}
\tablehead{
\colhead{UT Date} & \colhead{$\Delta$t$^b$} & \colhead{Q(\%)} & \colhead{U(\%)}  & 
\colhead{$\sigma$(Q,U)} & \colhead{P(\%)}  & \colhead{$\theta$(\arcdeg)} & 
\colhead{$\sigma(\theta)$}}
\startdata
April ~3$^c$ & 7.2~~ &\phs 0.492 & $-$0.123~ & 0.05~ & 0.507 & 173  & ~3  \nl
         & ~~~~~ & $-$0.109 &  0.0725 & 0.112 & 0.13~ & 73   & 24  \nl
April ~7 & 10.72 &\phs 0.221 & $-$0.018~ & 0.106 & 0.19~ & 178  & 17  \nl
         & ~~~~~ & $-$0.379 &  0.182~ & 0.14~ & 0.40~ & 77   & ~9  \nl
April 20$^d$ & 23.69 &\phs 0.260 &  0.503~ & 0.017 & 0.566 & 31.3 & ~1  \nl
         & ~~~~~ & $-$0.340 &  0.702~ & 0.101 & 0.77~ & 58   & ~4  \nl
April 26 & 29.62 &\phs 0.532 &  0.785~ & 0.008 & 0.948 & 28.0 & 0.3 \nl
         & ~~~~~ & $-$0.0673&  0.986~ & 0.10~ & 0.98~ & 47   & ~3  \nl
April 27 & 30.67 &\phs 0.488 &  0.753~ & 0.013 & 0.897 & 28.5 & 0.5 \nl
         & ~~~~~ & $-$0.111 &  0.954~ & 0.101 & 0.95~ & 48   & 3~  \nl
April 28 & 31.63 &\phs 0.557 &  0.668~ & 0.05~ & 0.870 & 25.1 & 1.7 \nl
         & ~~~~~ & $-$0.0425&  0.868~ & 0.112 & 0.86~ & 46   & ~4  \nl
April 30 & 33.70 &\phs 0.539 &  0.684~ & 0.006 & 0.871 & 26.0 & 0.2 \nl
         & ~~~~~ & $-$0.0605&  0.884~ & 0.10~ & 0.88~ & 47   & ~3  \nl
May   11 & 44.69 &\phs 0.411 &  0.492~ & 0.015 & 0.641 & 25.1 & 0.7 \nl
         & ~~~~~ & $-$0.189 &  0.691~ & 0.101 & 0.71~ & 53   & ~4  \nl
May   14 & 47.68 &\phs 0.447 &  0.545~ & 0.02~ & 0.70~ & 25.3 & ~1  \nl
         & ~~~~~ & $-$0.153 &  0.744~ & 0.102 & 0.75~ & 51   & ~4  \nl
\tablenotetext{a}{For each epoch, the first line shows the {\it observed}
polarization, and the second line shows the polarization {\it after} 
correction for ISpol. Except for April 7 which was a true $V$-band measurement,
results are flux-weighted averages of spectropolarimetry over 
the range 4800--6800 \AA, simulating similar bandpass coverage for 
comparison. The ISpol derived in this 
paper has been adopted. See text for details.} 
\tablenotetext{b}{Days since 1993 March 27.5 UT.}
\tablenotetext{c}{Average of April 3 and 4 measurements.}
\tablenotetext{d}{Averaged in the interval 6000--6800 \AA.}

\enddata
\end{deluxetable}

\clearpage
\begin{deluxetable}{llllll}
\scriptsize
\tablecolumns{6}
\tablewidth{0pc}
\tablecaption{Results of Fit to April 30 and May 11 Data$^a$}
\tablehead{
\colhead{} & \multicolumn{2}{c}{April 30}  & \colhead{}  &
\multicolumn{2}{c}{May 11} \\
\cline{2-3} \cline{5-6} \\
\colhead{Components} & \colhead{$Q$ (\%)} & \colhead{$U$ (\%)} & 
\colhead{}           & \colhead{$Q$ (\%)} & \colhead{$U$ (\%)}}
\startdata
Continuum          & \phs 0.89$\pm$0.05 &   \phs 0.74$\pm$0.05 & & 
                      \phs 0.76$\pm$0.10 &   \phs 0.26$\pm$0.09     \nl
H$\alpha$ emission &$-$0.39$\pm$0.11 & $-$0.04$\pm$0.11 & & 
                    $-$0.34$\pm$0.18 & \phs 0.48$\pm$0.18    \nl
Unshifted He I     &\phs 0.04$\pm$0.18\phs &\phs 0.44$\pm$0.18\phs & & 
                    \phs 0.81$\pm$0.24\phs &\phs 0.47$\pm$0.23\phs     \nl
Redshifted He I    &\phs 8.3\phn$\pm$1.1\phn & $-$2.2\phn$\pm$1.1\phn & & 
                    \phs 4.3\phn$\pm$2.0\phn & $-$1.9\phn$\pm$2.0\phn    \nl
\tablenotetext{a}{Fit performed over H$\alpha$ emission only, 6390--6860 \AA.}

\enddata
\end{deluxetable}

\clearpage
\begin{deluxetable}{lccccl}
\scriptsize
\tablecolumns{6}
\tablewidth{0pc}
\tablecaption{Continuum and H$\alpha$ Polarizations on April 20$^a$}
\tablehead{
\colhead{} & \multicolumn{2}{c}{Continuum}  & \colhead{}  &
\multicolumn{2}{c}{H$\alpha$} \\
\cline{2-3} \cline{5-6} \\
\colhead{} & \colhead{$Q$ (\%)} & \colhead{$U$ (\%)} & 
\colhead{} & \colhead{$Q$ (\%)} & \colhead{$U$ (\%)}}
\startdata
This paper: Continuum-sub$^b$        & 0.15 &  0.75 & & 0.60 & $-$0.20 \nl
~~~~~~~~~~~~~~~~~Spectral fitting$^b$ & 0.30 &  0.69 & & 0.40 & $-$0.30 \nl
                                     &      &       & &      &         \nl
THW                                  & 0.33 &  0.63 & & 0.55 & $-$0.95 \nl
\tablenotetext{a}{Averaged over spectral region containing H$\alpha$ emission, 
6390--6860 \AA. Uncertainty is $\sim$ $\pm$0.1\%.}
\tablenotetext{b}{``Continuum-sub" is the method of obtaining line and
continuum polarizations by subtraction of the continuum in Stokes and total 
fluxes. ``Spectral fitting" achieves the same by least-squares fitting of 
the Stokes fluxes. See Appendix A for discussion.}

\enddata
\end{deluxetable}

\clearpage
\centerline{\rm FIGURE CAPTIONS}

\figcaption[]{A montage of the total flux spectra of SN 1993J for which
spectropolarimetry was obtained. Except for the top (April 20) curve, 
all spectra have been shifted vertically for clarity.
The April 27 spectrum, which is similar to that of April 26, is not shown.}

\figcaption[]{Observed spectropolarimetry of SN 1993J on 1993 April 20.
From top to bottom are the (a) total flux spectrum, (b) observed 
degree of polarization, (c) polarization position angle, and (d) 
polarized flux spectrum. Note the drop of \p~and rotation of PA across 
the P-Cygni profile of H$\alpha$. In (a) and elsewhere in this paper, 
telluric absorption is present near 6900 \AA, 7200 \AA, and 
6290 \AA~(very small dip).}

\figcaption[]{Spectropolarimetry of SN 1993J on 1993 April 20, displayed
as in Figure 2 but after correction for the ISpol value derived in this paper. 
Note the disappearance of rotation in PA across H$\alpha$ and little
or no H$\alpha$ emission in the \pf~spectrum. However, a ``polarized feature", 
absent in Figure 2, has appeared near 4650 \AA.}

\figcaption[]{The (a) polarization and (b) polarized flux spectra of 
April 20, after correction for the ISpol of THW. Note that in comparison with 
Figure 3, the polarization is significantly higher and there is still 
substantial H$\alpha$ residual emission.}

\figcaption[]{Observed spectropolarimetry of SN 1993J on 1993 April 30, arranged
as in Figure 2.}

\figcaption[]{Observed spectropolarimetry of SN 1993J on 1993 May 11, arranged
as in Figure 2.}

\figcaption[]{The top panel (a) illustrates the observed ({\it thin line}) 
and model fitted ({\it heavy line}) flux spectra of SN 1993J on April 30. 
Also shown in the lower portion is the H-rich line component used in the fit. 
Panel (b) shows the redshifted He-rich line component derived from the 
spectrum of the prototypical Type Ib SN 1984L (Harkness \etal~1987). 
In (c) the observed 
({\it thin line}) and fitted ({\it thick line}) \qf~spectra are shown.  
Note the remarkable overall resemblance of \qf~to the redshifted SN~1984L 
spectrum. For clarity, a third component used in the fit, the unshifted 
spectrum of SN~1984L, is not shown.}

\figcaption[]{The results of the fit to (a) the observed total flux \flam,
(b) \qf, and (c) \uf~spectra of April 30. The thin line shows the observed 
spectrum and the thick line shows the fitted curve around H$\alpha$ emission 
only. A horizontal line drawn under H$\alpha$ in the top panel denotes the 
adopted continuum. Note that here, the fit gives better agreement
around H$\alpha$ emission, where it was constrained, than in Figure 7, 
where it was performed over a larger wavelength range. The relatively
poor fit around 6400--6500 \AA~may be due to blueshifted absorption of
polarized light by He and H gas.}

\figcaption[]{As in Figure 8, but for the May 11 data.}

\figcaption[]{Observed and ISpol-corrected polarizations of SN 1993J 
(from Table 2) plotted in ($Q,U$) space. Open circles are {\it observed} 
quantities; solid circles are values after correction for ISpol. Numbers
1 and 9 correspond to the first and last dates of our measurements, 
respectively. The asterisk denotes the ISpol derived in this paper and 
the triangle shows the THW ISpol. Note that our ISpol correction nicely 
shifts the points so that the very first early measurement gives an 
essentially null polarization for the supernova intrinsically.} 

\figcaption[]{(a) The $V$-band light curve of SN 1993J, from 
Lewis \etal~(1994), (b) $V$-band ISpol-corrected polarization, 
and (c) position angle as a function of time. 
The polarization is low at early times but rapidly rises to a maximum 
and then declines, while being accompanied by a rotation in PA.}

\figcaption[]{Spectropolarimetry of SN 1993J on 1993 April 30 after correction
for the ISpol value derived in this paper. Arranged as in Figure 2.}

\figcaption[]{As in Figure 12, but using the ISpol value derived by THW. 
Note that the shape, redshift, and strength of various lines in the 
\pf~spectrum appear to be different from those in Figure 12.}

\figcaption[]{Spectral fitting to determine continuum and H$\alpha$ 
polarizations on April 20. 
(a) The observed flux spectrum ({\it top, thin line}), overlayed with the 
best model ({\it heavy line}) obtained by combining a constant continuum with
the line emission component from the April 15 spectrum ({\it lower curve}). 
The horizontal line denotes the constant continuum level and extent of fit.
(b), (c) The observed Stokes flux spectra \qf~and \uf, respectively, with
the best fit shown as heavy lines.} 

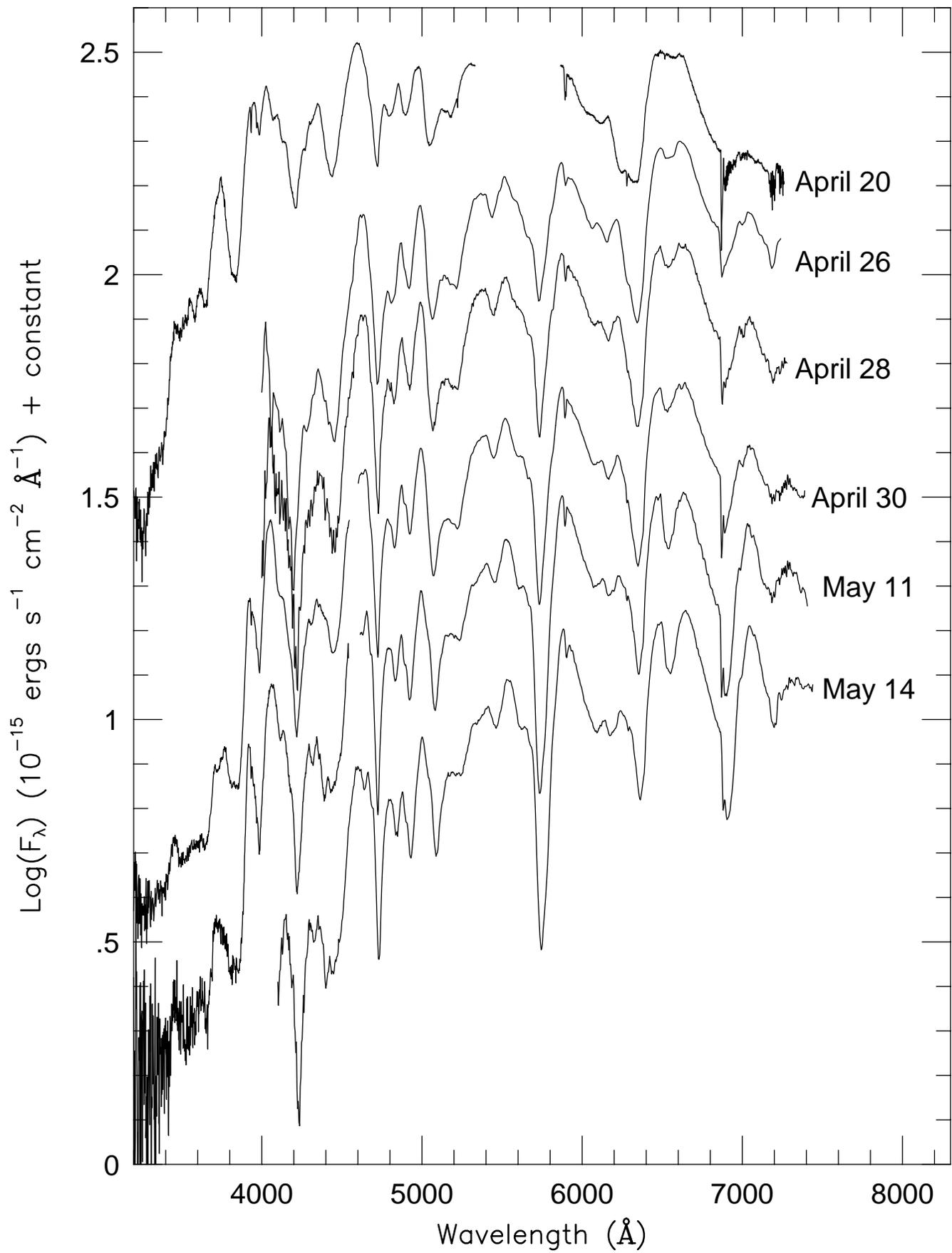
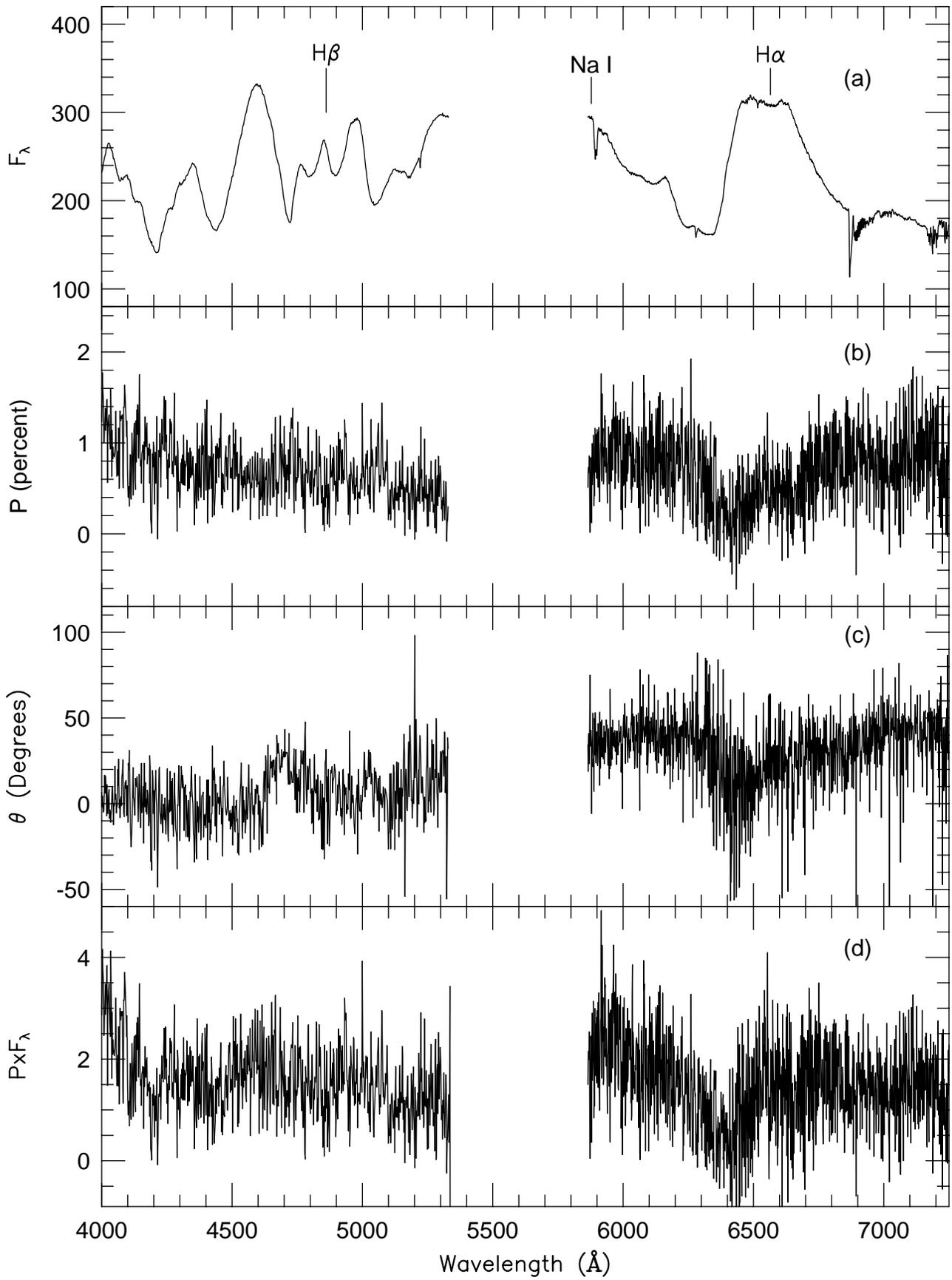
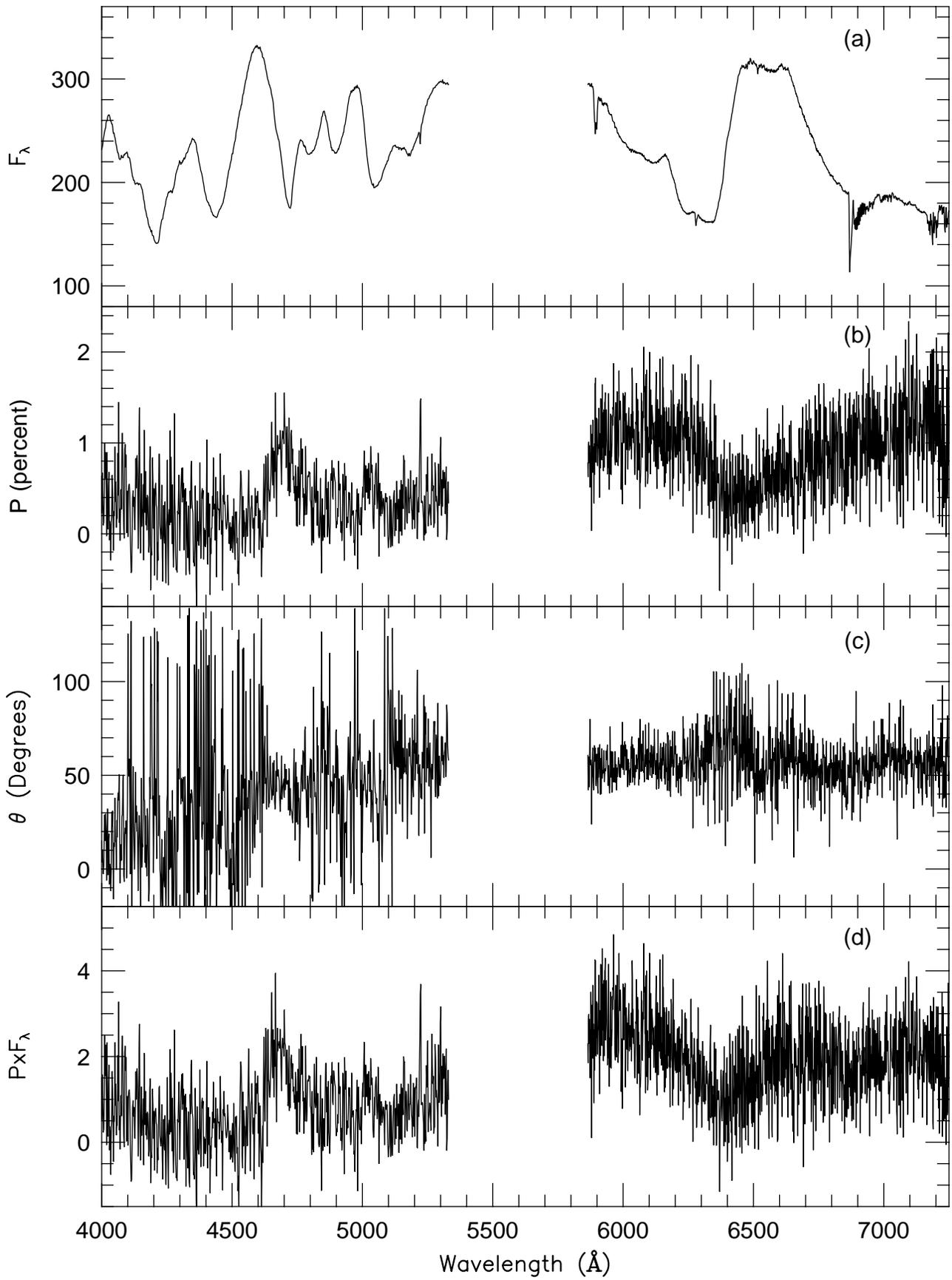
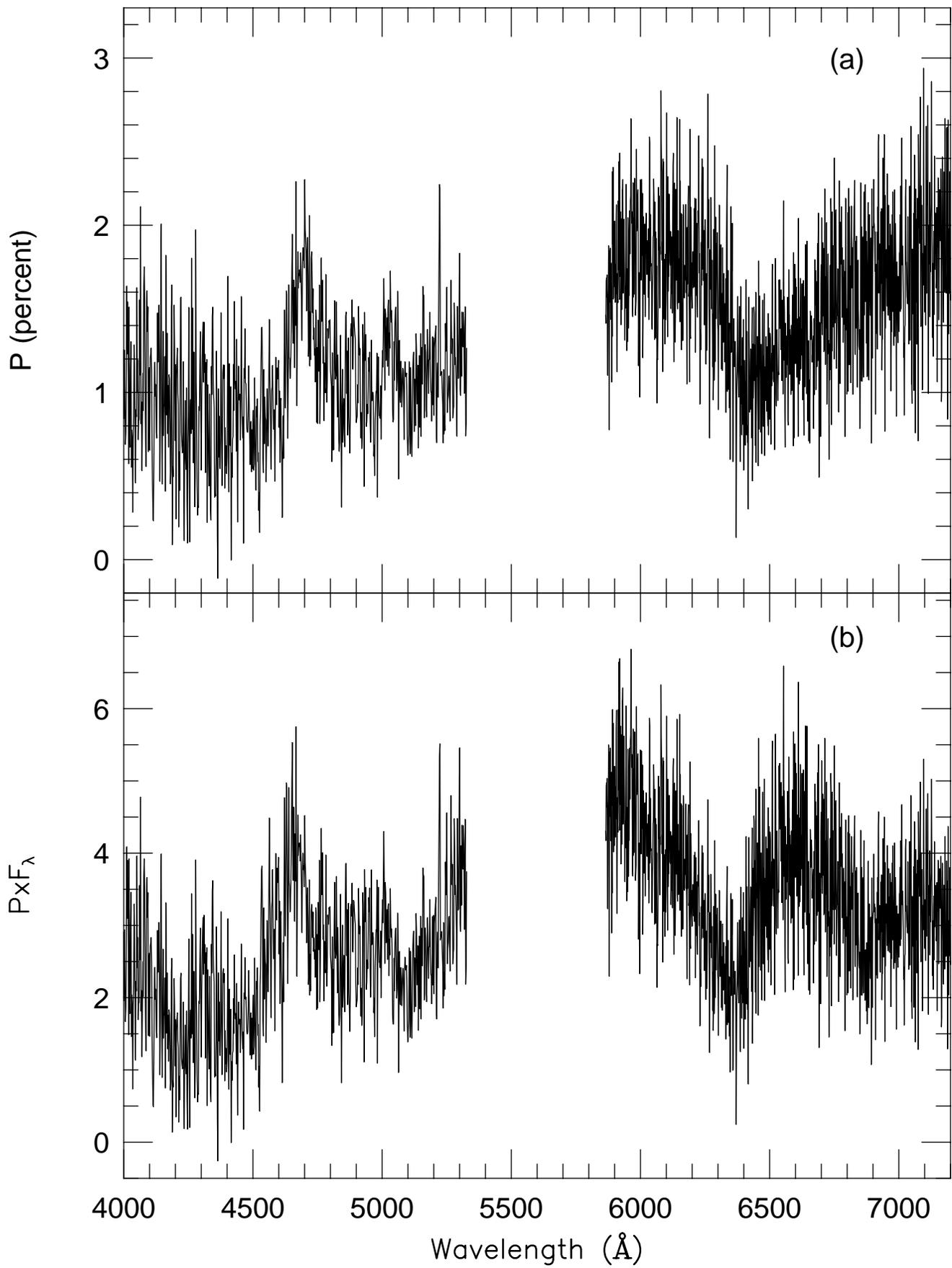
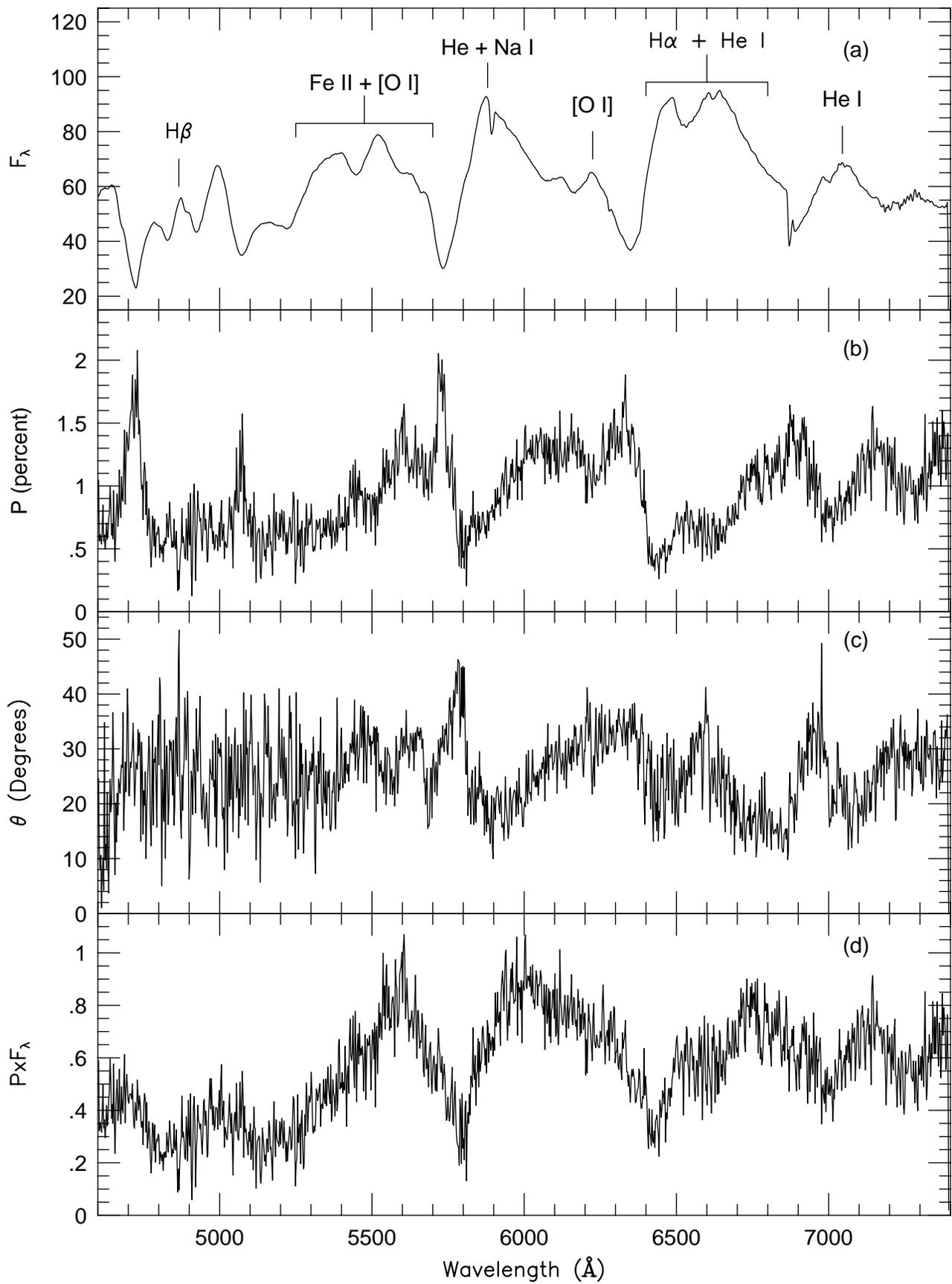
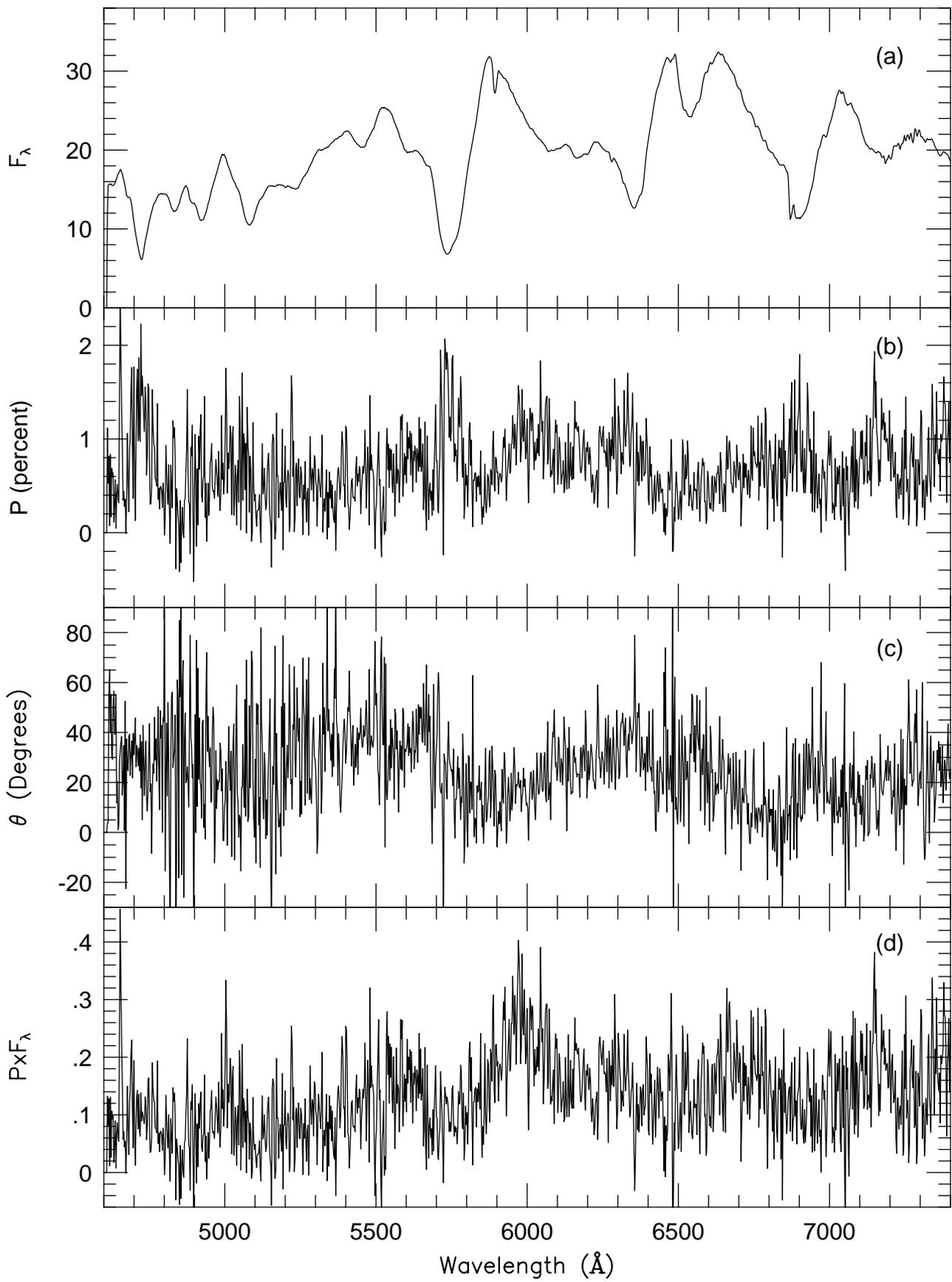
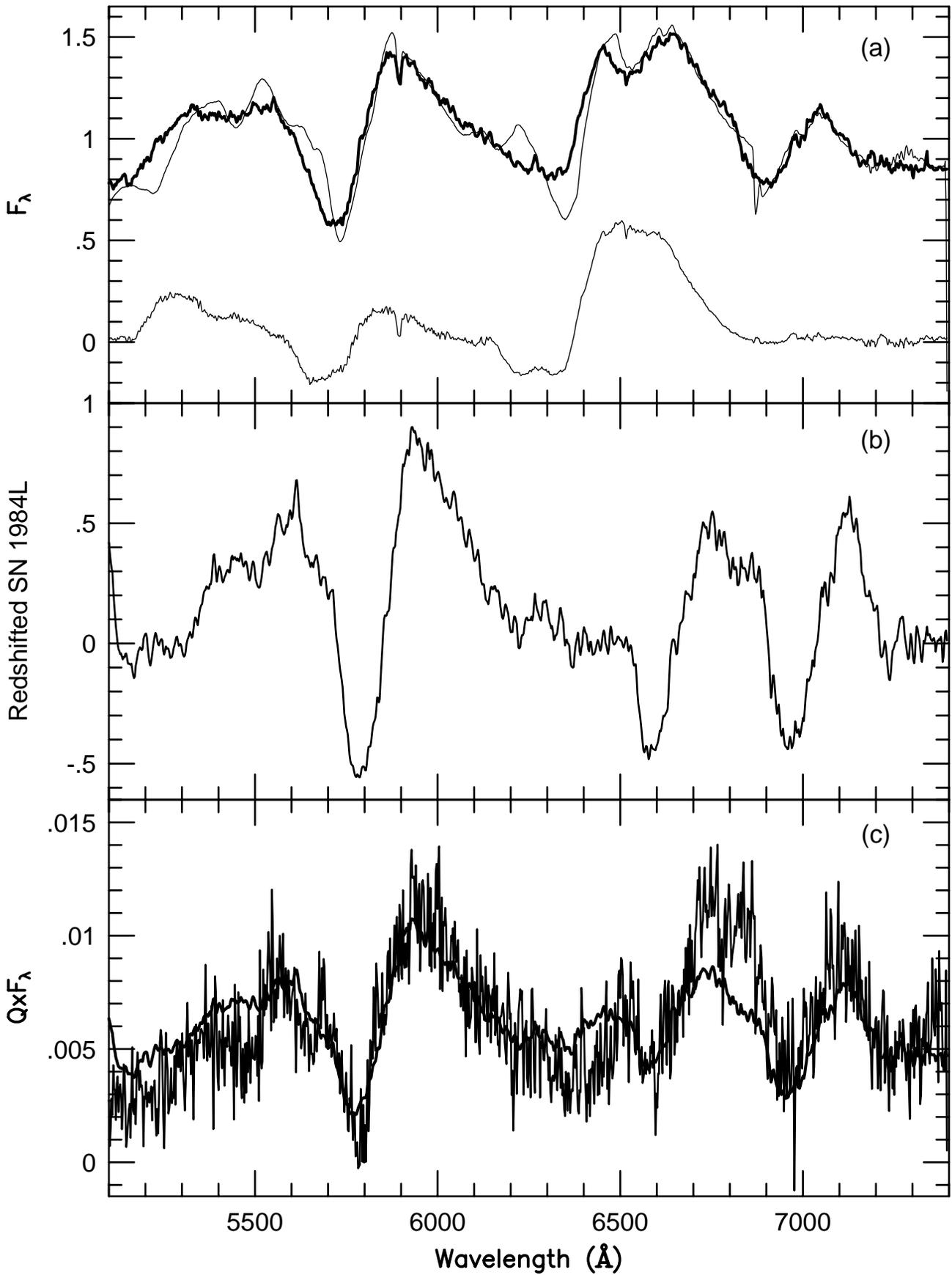
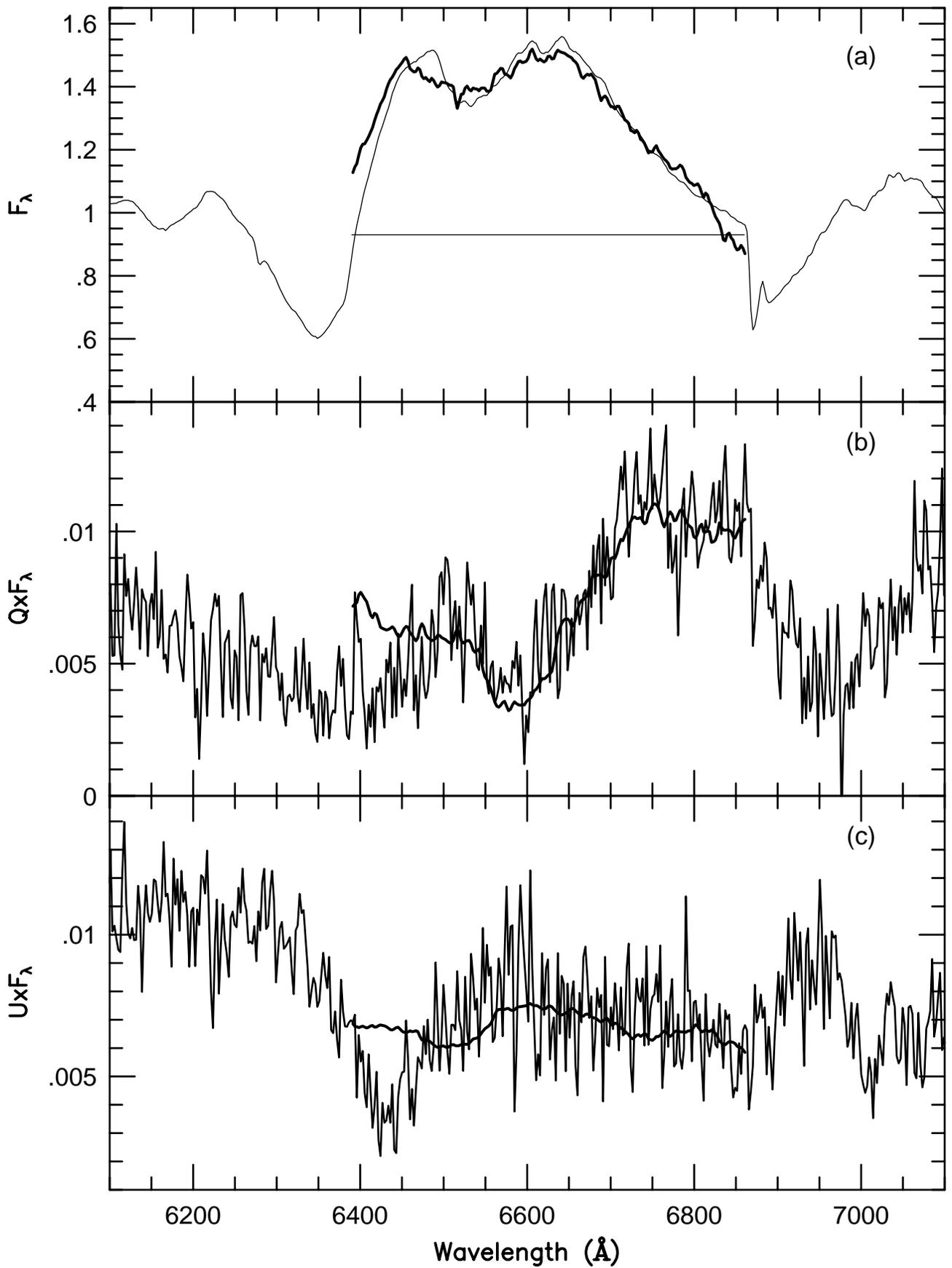
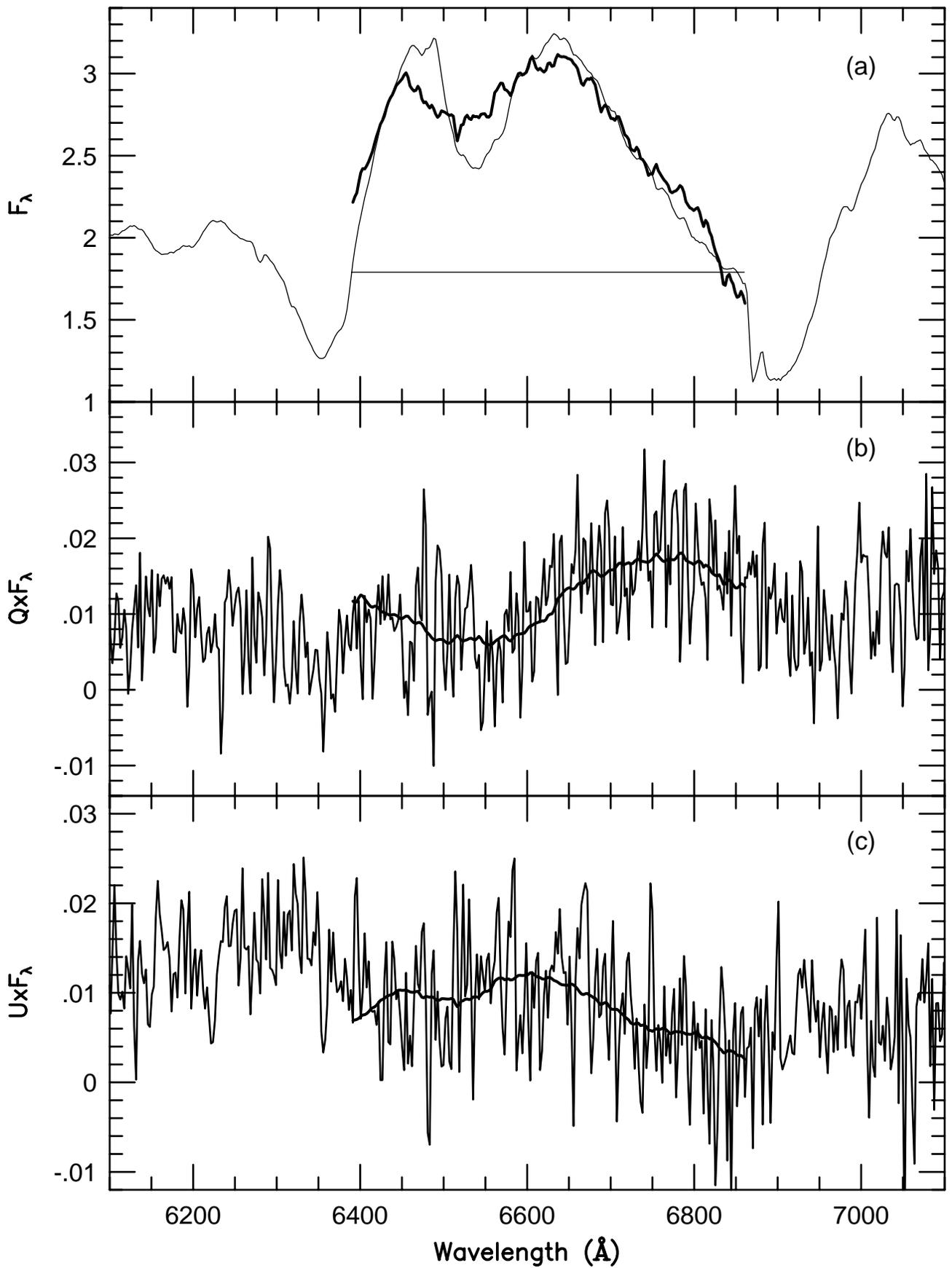
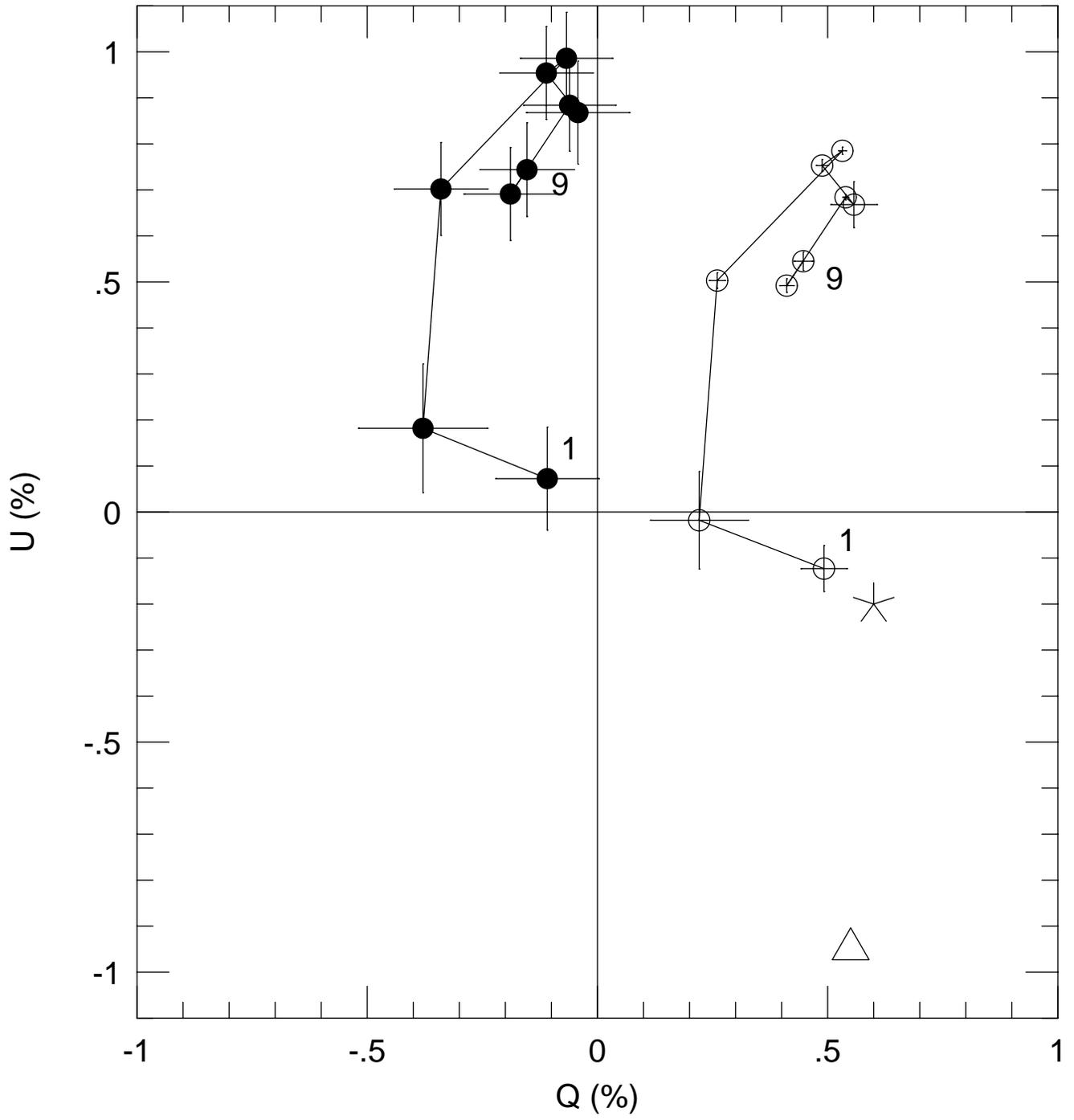
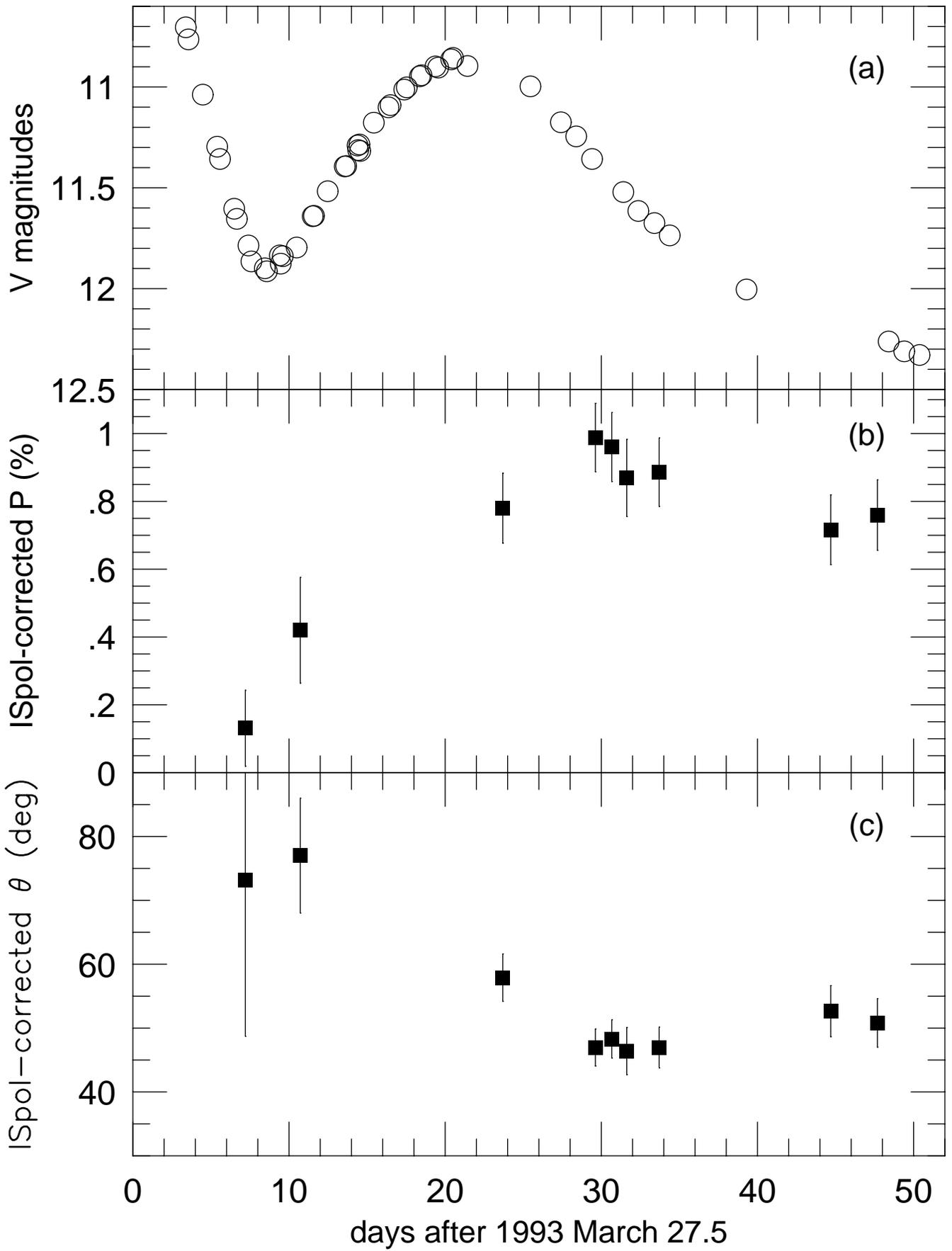
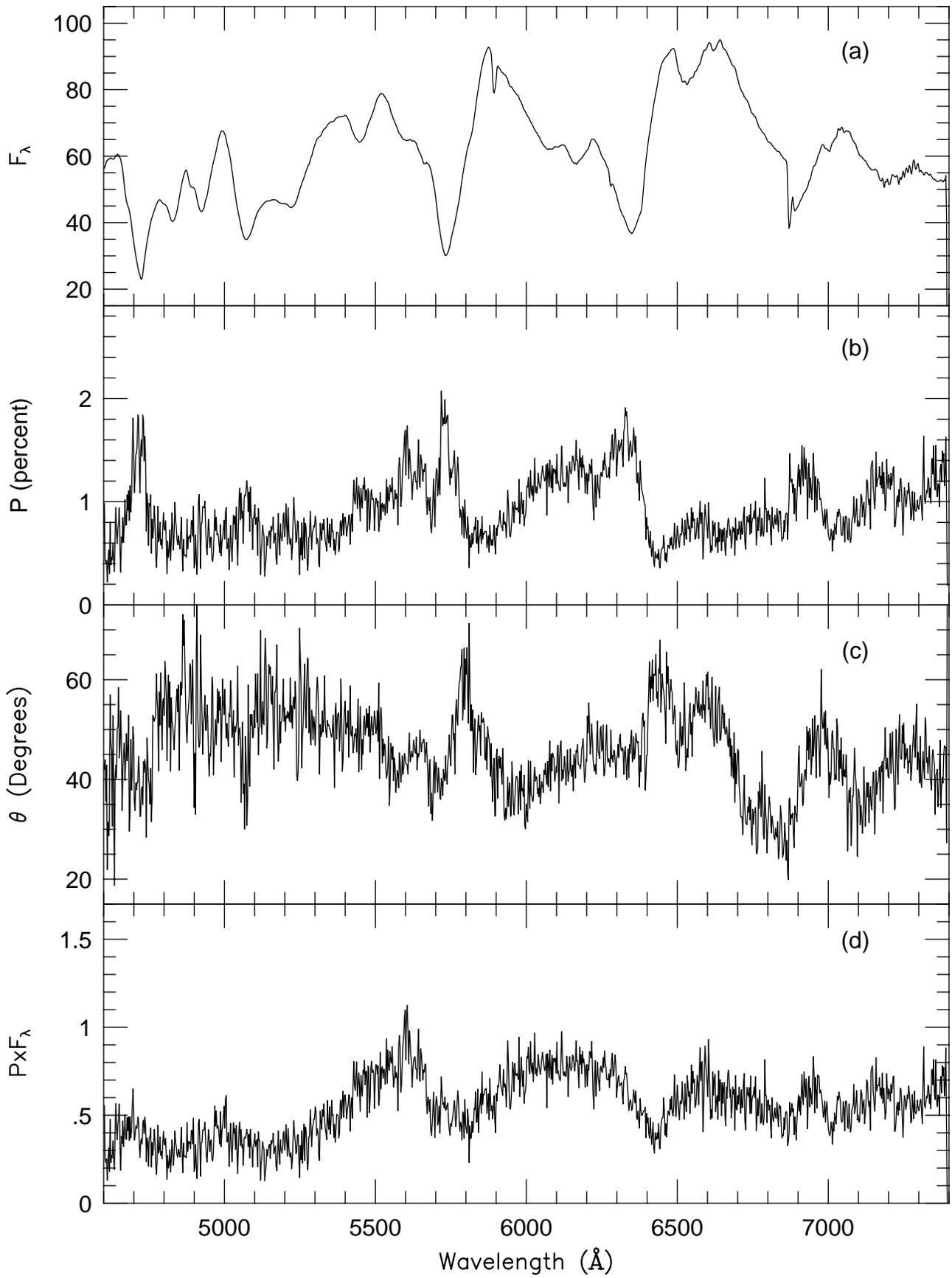
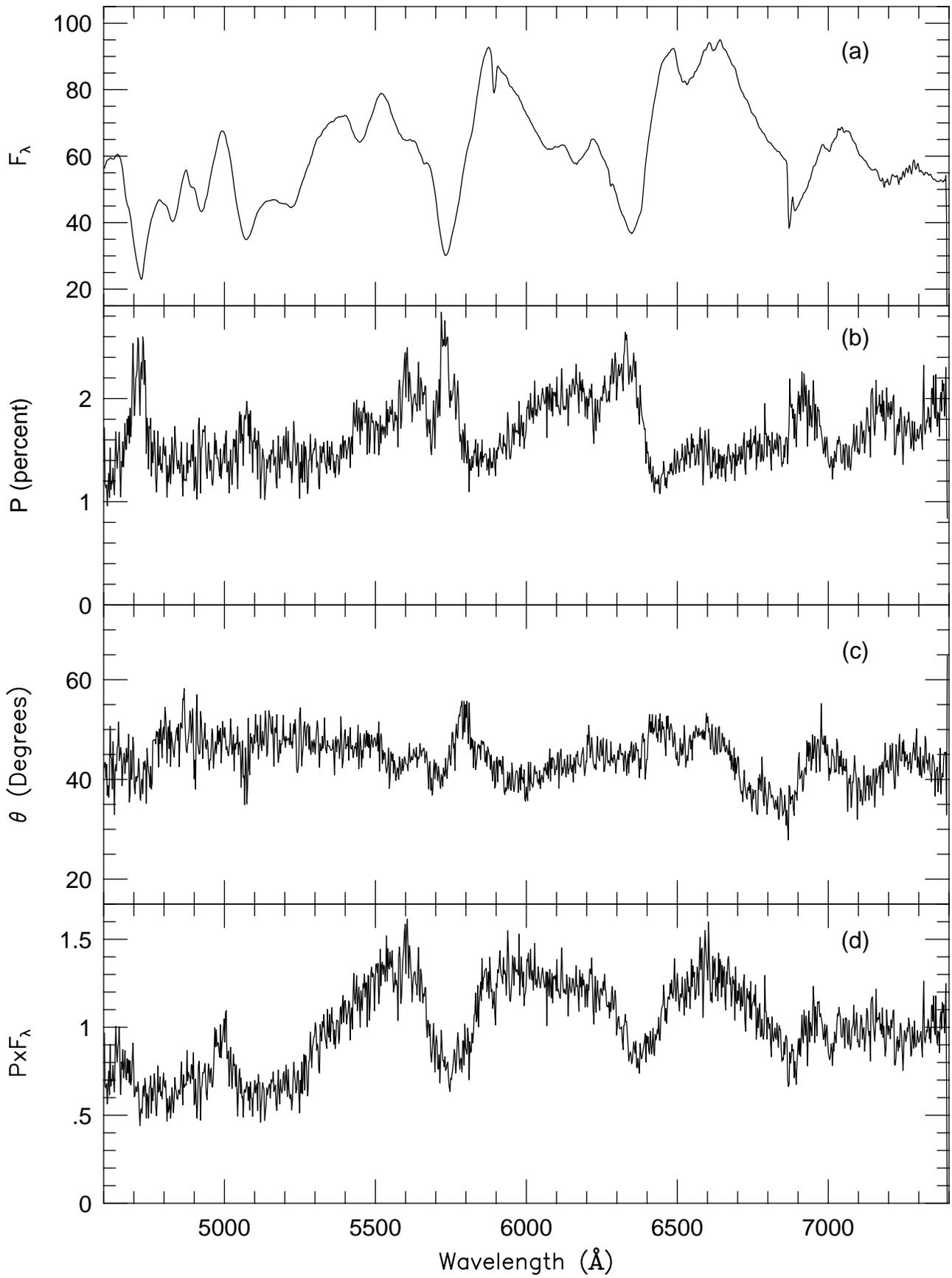
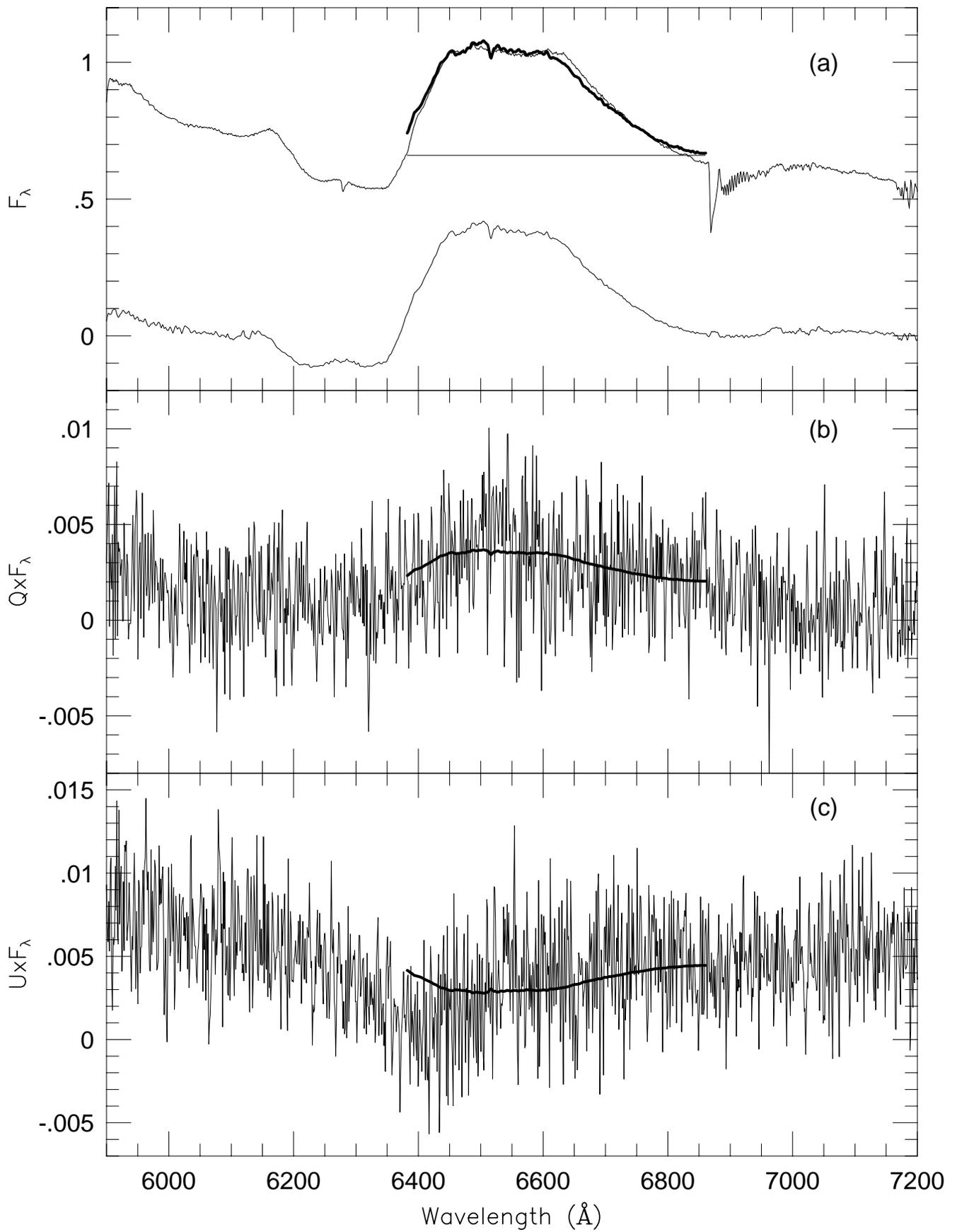
\end{document}